\documentclass[multphys,vecphys]{svmult}
\usepackage{graphicx}
\usepackage{multicol}
\usepackage{amssymb}

\newcommand{\R}{{\mathbb R}}
\newcommand{\lP}{\ell_{\rm P}}
\newcommand{\sgn}{\mathop{\mathrm{sgn}}\nolimits}
\newcommand{\tr}{\mathop{\mathrm{tr}}\nolimits}
\newcommand{\be}{\begin{equation}}
\newcommand{\ee}{\end{equation}}
\newcommand{\ba}{\begin{eqnarray}}
\newcommand{\ea}{\end{eqnarray}}

\begin{document}

\title*{Cosmological applications of loop quantum gravity\protect\footnote{Preprint CGPG--03/6--1}}
\titlerunning{Loop quantum gravity}
\author{Martin Bojowald\inst{1}\and
Hugo A. Morales-T\'ecotl\inst{2}\inst{3}}
\institute{Center for Gravitational Physics and Geometry, The
  Pennsylvania State University, University Park, PA 16802, USA
\texttt{bojowald@gravity.psu.edu}
\and Departamento de F\'{\i}sica, Universidad Aut\'onoma Metropolitana Iztapalapa,
A.P. 55-534 M\'exico D.F. 09340, M\'exico \texttt{hugo@xanum.uam.mx}\and
Associate member of AS-ICTP Trieste, Italy.
}

\maketitle

\section{Introduction}
\label{s:Intro}

According to general relativity, not only the gravitational field but
also the structure of space and time, the stage for all the other
fields, is governed by the dynamical laws of physics. The space we see
is not a fixed background, but it evolves on large time scales, even
to such extreme situations as singularities where all of space
collapses into a single point. At such a point, however, energy
densities and tidal forces diverge; all classical theories break down,
even general relativity itself. This implies that general relativity
cannot be complete since it predicts its own breakdown. Already for a
long time, it has been widely expected that a quantum theory of
general relativity would cure this problem, providing a theory which
can tell us about the fate of a classical singularity.

Most of the time, quantum gravity has been regarded as being far away
from observational tests. In such a situation, different approaches
would have to be judged purely on grounds of internal consistency and
their ability to solve conceptual problems. Those requirements are
already very restrictive for the quantization of a complicated theory
as general relativity, to the extent that in all the decades of
intense research not a single completely convincing quantum theory of
gravity has emerged yet, even though there is a number of promising
candidates with different strengths. Still, the ultimate test of a
physical theory must come from a confrontation with observations of
the real world. For quantum gravity, this means observations of
effects which happen at the smallest scales, the size of the Planck
length $\lP\approx 10^{-32}{\rm cm}$.

In particular in the light of recent improvements in precision
cosmology, the cosmological arena seems to be most promising for
experimental tests. This is fortunate since also many conceptual
issues arise in the cosmological setting where the universe is studied
as a whole. Examples are the singularity problem mentioned above and
the so-called problem of time which we will address later. Therefore,
one can use the same methods and approximations to deal with
conceptual problems and to derive observational consequences.

One of the main approaches to quantum gravity is based on a canonical
quantization of general relativity, which started with the formal
Wheeler--DeWitt quantization and more recently evolved into quantum
geometry. Its main strength is its background independence, i.e.\ the
metric tensor which describes the geometry of space is quantized as a
whole and not split into a background and a dynamical part. Since most
familiar techniques of quantum field theory rely on the presence of a
background, for this ambitious approach new techniques had to be
invented which are often mathematically involved. By now, most of the
necessary methods have been developed and we are ready to explore
them in simple but physically interesting situations.

Being the quantization of a complicated, non-linear field theory,
quantum gravity cannot be expected to be easily understood in full
generality. As always in physics, one has to employ approximation
techniques which isolate a small number of objects one is interested
in without taking into account all possible interactions. Prominent
examples are symmetric models (which are usually called
minisuperspaces in the context of general relativity) and
perturbations of some degrees of freedom around a simple
solution. This opens the possibility to study the universe as a whole
(which is homogeneous and isotropic at large scales) as well as the
propagation of a single particle in otherwise empty space (where
complicated interactions can be ignored).

These two scenarios constitute the two main parts of this article. In
the context of the first one (Section \ref{s:Cosmo}) we discuss the
basic equations which govern the quantum evolution of an isotropic
universe and special properties which reflect general issues of
quantum gravity. We then analyze these equations and see how the
effects of quantum geometry solve and elucidate important conceptual
problems. Quantum gravity effects in these regimes also lead to
modifications of classical equations of motion which can be used in a
phenomenological analysis. A different kind of phenomenology, related
to the propagation of particles in empty space, is discussed in
Section \ref{s:Phen}. In this context cosmological scales
are involved for many proposals of observations, and so they fit into the present scheme.

Both settings are now at a stage where characteristic effects have
been identified and separated from the complicated, often intimidating
technical foundation. This is a natural starting point for
phenomenological analyzes and opens a convenient port of entry for
beginners to the field.

The article is intended to describe the basic formalism to an extent
which makes it possible to understand the applications without
requiring too much background knowledge (the presentation cannot be
entirely background independent, though). The general framework of
quantum geometry is reviewed briefly in Section \ref{s:QG} after
recalling facts about general relativity (Section \ref{s:GR}) and the
Wheeler--DeWitt quantization (Section \ref{s:WdW}). For the details we
provide a guide to the literature including technical reviews and
original papers.

\section{General relativity}
\label{s:GR}

General relativity is a field theory for the metric $g_{\mu\nu}$ on a
space-time $M$ which determines the line element\footnote{In
expressions with repeated indices, a summation over the allowed range
is understood unless specified otherwise. We use greek letters
$\mu,\nu,\ldots$ for space-time indices ranging from zero to three and
latin letters from the beginning of the alphabet, $a,b,\ldots$ for
space indices ranging from one to three.}
$\D s^2=g_{\mu\nu}(x)\D x^{\mu}\D x^{\nu}$. The line element, in turn
specifies the geometry of space-time; we can, e.g., measure the length
of a curve $C\colon\R\to M, t\mapsto x^{\mu}(t)$ using
\[
 \ell(C)=\int \D s = \int \sqrt{g_{\mu\nu}(x(t))\dot{x}^{\mu}(t)
 \dot{x}^{\nu}(t)}\, \D t\,.
\]

\subsection{Field equations}

While a space-time can be equipped with many different metrics,
resulting in different geometries, only a subclass is selected by
Einstein's field equations of general relativity which are complicated
non-linear partial differential equations with the energy density of
matter as a source.

They can be understood as giving the dynamical evolution of a
space-like geometry in a physical universe. Due to the four-dimensional
covariance, however, in general there is no distinguished space-like
slice which could be used to describe the evolution. All possible slices
are allowed, and they describe the same four-dimensional picture
thanks to symmetries of the field equations.

Selecting a slicing into space-like manifolds, the field equations
take on different forms and do not show the four-dimensional
covariance explicitly. However, such a formulation has the advantage
that it allows a canonical formulation where the metric $q_{ab}$ only
of space-like slices plays the role of coordinates of a phase space,
whose momenta are related to the time derivative of the metric, or the
extrinsic curvature $K_{ab}=-\frac{1}{2}\dot{q}_{ab}$ of a slice
\cite{ADM}. This is in particular helpful for a quantization since
canonical quantization techniques become available. The momentum
conjugate to the metric $q_{ab}$ is related to the extrinsic curvature
by 
\[
 \pi^{ab}= -{\textstyle\frac{1}{2}}\sqrt{\det q}\,(K^{ab}-q^{ab}K^c_c)
\]
where indices are raised by using the inverse $q^{ab}$ of the metric.
The dynamical field equation, the analog of Einstein's field
equations, takes the form of a constraint,\footnote{Note that this
requires a relation between the basic fields in every point of space;
there are infinitely many degrees of freedom and infinitely many
constraints.} the Hamiltonian constraint
\begin{eqnarray} \label{HADM}
 && 8\pi G\sqrt{\det q}\,(q_{ac}q_{bd}+q_{ad}q_{bc}-q_{ab}q_{cd})
  \pi^{ab}\pi^{cd} -\frac{1}{16\pi G}\sqrt{\det q}
  \;\;\mbox{}^3\!R(q)\nonumber\\
  &&+\sqrt{\det q}\;\rho_{\rm matter}(q)=0
\end{eqnarray}
where $G$ is the gravitational constant, $\mbox{}^3\!R(q)$ the so-called
Ricci scalar of the spatial geometry (which is a function of the
metric $q$), and $\rho_{\rm matter}(q)$ is the energy density of matter
depending on the particular matter content (it depends on the metric,
but not on its momenta in the absence of curvature couplings).

The complicated constraint can be simplified slightly by transforming
to new variables \cite{AshVar}, which has the additional advantage of
bringing general relativity into the form of a gauge theory, allowing
even more powerful mathematical techniques.  In this reformulation,
the canonical degrees of freedom are a densitized triad $E^a_i$ which
can be thought of as giving three vectors labelled by the index $1\leq
i \leq 3$. Requiring that these vectors are orthonormal defines a
metric given by
\[
 q_{ab}=\sqrt{|\det E^c_j|}(E^{-1})_a^i (E^{-1})_b^i\,.
\]
Its canonical conjugate is the Ashtekar connection
\begin{equation}
 A_a^i= \Gamma_a^i-\gamma K_a^i
\end{equation}
where $\Gamma_a^i$ is the spin connection (given uniquely by the triad
such that $\partial_a E^b_i+ \epsilon_{ijk}\Gamma^j_a E^b_k=0$) and
$K_a^i$ is the extrinsic curvature. The positive Barbero--Immirzi
parameter $\gamma$ also appears in the symplectic structure together
with the gravitational constant $G$
\begin{equation} \label{Poisson}
 \{A_a^i(x),E^b_j(y)\}=8\pi\gamma G\delta_a^b\delta_j^i\delta(x,y)
\end{equation}
and labels equivalent classical formulations. Thus, it can be chosen
arbitrarily, but the freedom will be important later for the quantum
theory. The basic variables can be thought of as a ``vector
potential'' $A_a^i$ and the ``electric field'' $E^a_i$ of a gauge
theory, whose gauge group is the rotation group SO(3) which rotates
the three triad vectors: $E^a_i\mapsto \Lambda^j_i E^a_j$ for
$\Lambda\in {\rm SO(3)}$ (such a rotation does not change the metric).

Now, the Hamiltonian constraint takes the form \cite{AshVarReell}:
\begin{eqnarray} \label{HAsh}
 &&|\det E_l^c|^{-1/2}\epsilon_{ijk}
  F^i_{ab}E^a_jE^b_k-2(1+\gamma^2)|\det E_l^c|^{-1/2}
  K_{[a}^iK_{b]}^j E_i^aE_b^j\nonumber\\
 && +8\pi G \sqrt{|\det E^c_l|}\; \rho_{\rm
  matter}(E)=0
\end{eqnarray}
where $F_{ab}^i$ the
curvature of the Ashtekar connection, and the matter energy density
$\rho_{\rm matter}(E)$ now depends on the triad via the metric.

\subsection{Approximations}

Given the complicated nature of the field equations, one has to resort
to approximation schemes in order to study realistic situations. In
the case of gravity, the most widely used approximations are:
\begin{itemize}
\item Assume symmetries. This simplifies the field equations by
 eliminating several degrees of freedom and simplifying the relations
 between the remaining ones. In a cosmological situation, for
 instance, one can assume space to be homogeneous such that the field
 equations reduce to ordinary differential equations in time.
\item
 Perturbations around a simple known solution. One can, e.g., study a
 small amount of matter, e.g.\ a gravitational wave or a single
 particle, and its propagation in Minkowski space. To leading order,
 the back reaction of the geometry, which changes due to the presence
 of the particle's energy density, on the particle's propagation can
 be ignored.
\item Asymptotic regimes with boundary conditions. In
 many situations it is possible to isolate interesting degrees of
 freedom by looking at boundaries of space-time with special boundary
 conditions capturing the physical situation. It can then be possible
 to ignore interactions with the bulk degrees of freedom which
 simplifies the analysis. This strategy is most widely used in the
 context of black hole physics, in its most advanced form with
 isolated horizon conditions; see, e.g.\ \cite{IH}.
\end{itemize}

The first two approximation schemes and their applications in quantum
geometry will be discussed on Sections \ref{s:Cosmo} and \ref{s:Phen},
respectively. Since the last one so far does not have many
cosmological applications, it will not be used here. It does have
applications in quantum geometry, however, in the calculation of black
hole entropy \cite{Entro}. In this section we only
illustrate the first one in the context of isotropic cosmology.

\subsection{Cosmology}

In the simplest case of a cosmological model we can assume space to be
isotropic (looking the same in all its points and in all directions)
which implies that one can choose coordinates in which the line
element takes the form
\begin{equation}\label{dsflat}
 \D s^2=-\D t^2+a(t)^2((1-kr^2)\D
 r^2+r^2(\D\vartheta^2+\sin\vartheta\D\varphi^2))
\end{equation}
with the scale factor $a(t)$ (the evolving ``radius'' of the
universe). The constant $k$ can take the values $k=0$ for a spatially
flat model (planar), $k=1$ for a model with positive spatial curvature
(spherical), and $k=-1$ for a model with negative spatial curvature
(hyperbolic). Einstein's field equations restrict the possible
behavior of $a(t)$ in the form of the Friedmann equation
\cite{Friedmann}
\begin{equation} \label{Friedmann}
 \left(\frac{\dot{a}}{a}\right)^2=\frac{16\pi}{3}G\,\rho(a)-\frac{k}{a^2}\,.
\end{equation}
Since also the matter density $\rho(a)$ enters, we can find $a(t)$
only if we specify the matter content. Common choices are ``dust''
with $\rho(a)\propto a^{-3}$ or ``radiation'' with $\rho(a)\propto
a^{-4}$ (due to an additional red-shift factor), which describe the
matter degrees of freedom collectively. After choosing the matter
content, we just need to solve an ordinary differential equation. For
radiation in a spatially flat universe, e.g., all solutions are given
by $a(t)\propto\sqrt{t-t_0}$ where $t_0$ is an integration constant.

In a more complicated but also more fundamental way one can describe
the matter by using additional matter fields\footnote{In a homogeneous
model, matter ``fields'' are also described by a finite number of
parameters only, e.g.\ a single one for a scalar $\phi$.} which enter
via their Hamiltonian (or total energy). This results in a system of
coupled ordinary differential equations, one for the scale factor and
others for the matter fields. A common example in cosmology is a
scalar $\phi$ which has Hamiltonian
\begin{equation} \label{Hphi}
 H_{\phi}(a)= {\textstyle\frac{1}{2}}a^{-3}p_{\phi}^2+a^3 W(\phi)
\end{equation}
with its potential $W$ and the scalar momentum
$p_{\phi}=a^3\dot{\phi}$. Note that it is important to keep track of
the $a$-dependence in cosmology since $a$ is evolving; in the usual
formulas for Hamiltonians on Minkowski space $a$ does not appear.

The Friedmann equation is now given by (\ref{Friedmann}) with energy
density $\rho(a)=H_{\phi}(a)/a^3$. Now, the right hand side depends
explicitly on $\phi$ and $p_{\phi}$ which both depend on time. Their
evolution is given by the Hamiltonian equations of motion
\begin{eqnarray}
 \dot{\phi} &=& \{\phi,H_{\phi}\}=p_{\phi}/a^3 \\
 \dot{p}_{\phi} &=& \{p_{\phi},H_{\phi}\} = -a^3 W'(\phi)\;.
\end{eqnarray}
By using the first equation one can transform the second one into a
second order equation of motion for $\phi$:
\begin{equation} \label{phidd}
 \ddot{\phi}=-3\,\dot{a}a^{-1}\, \dot{\phi}- W'(\phi)
\end{equation}
which in addition to the usual force term from the potential has a
friction term proportional to the first derivative of $\phi$. The
friction is strongest for a rapid expansion.

When we come close to $a=0$, the kinetic term usually dominates and
even diverges when $a=0$. This is problematic and leads to the
singularity problem discussed in the following subsection. However,
the divergence occurs only when $p_{\phi}\not=0$ for small $a$, so one
could try to arrange the evolution of the scalar such that the divergence is avoided. In
addition to suppressing the diverging kinetic term, we have the
additional welcome fact that $p_{\phi}\approx 0$ implies
$\phi\approx\phi_0={\rm const}$. The right hand side of the Friedmann
equation then becomes constant, $(\dot{a}/a)^2\approx\Lambda=(16\pi
G/3) W(\phi_0)$ for $k=0$. Its solutions are given by $a(t)\propto
\exp(\sqrt{\Lambda}t)$ which describes an {\em accelerated\/}
expansion, or inflation. Though motivated in a different way here,
inflation is deemed to be an important ingredient in cosmological
model building, in particular for structure formation. 

Unfortunately, however, it is very difficult to arrange the evolution
of the scalar in the way described here; for -- in addition to
introducing a new field, the inflaton $\phi$ -- it requires very
special scalar potentials and also initial values of the scalar. A
common choice is a quadratic potential $W(\phi)=\frac{1}{2}m\phi^2$
(e.g., for chaotic inflation) which requires a very small $m$ (a very
flat potential) for inflation to take place long enough, and also a
huge initial value $\phi_0$ pushing it up to Planck values. There is a
plethora of models with intricate potentials, all requiring very
special choices.

Inflation in general is the term for accelerated expansion \cite{GenInfl}, i.e.\
$\ddot{a}>0$. It is not necessarily of the exponential form as above, but
can be parameterized by different ranges of the so-called equation of
state parameter $w$ which needs to be less than $w<-\frac{1}{3}$ for
inflation. It can be introduced by a phenomenological $a$-dependence of
the energy density,
\begin{equation} \label{rhow}
 \rho(a)\propto a^{-3(w+1)}\;.
\end{equation}
Note, however, that this is in general possible only with
$a$-dependent $w$ except for special cases. Solutions for $a$ (with
$k=0$) are then of the form
\begin{equation} \label{inflation}
 a(t) \propto \left\{\begin{array}{clc} (t-t_0)^{2/(3w+3)} & \mbox{ for }
 -1<w<-\frac{1}{3} & \mbox{(power-law inflation)} \\
 \exp(\sqrt{\Lambda}t) & \mbox{ for } 
 w=-1  & \mbox{(standard inflation)} \\
 (t_0-t)^{2/(3w+3)} & \mbox{ for }
 w<-1 & \mbox{(super-inflation)} \end{array}\right.
\end{equation}
where $t_0$ is an initial value (replaced by $\Lambda=
(16\pi/3)G\rho$ with the constant energy
density $\rho$ for standard inflation). Note in particular that
super-inflation (also called pole-law inflation) can be valid only
during a limited period of time since otherwise $a$ would diverge for
$t=t_0$. While these possibilities add more choices for model
building, they share with standard inflation that they are difficult
to arrange with scalar potentials.

\subsection{Singularities}

Trying to suppress the kinetic term has led us to introduce inflation
as an ingredient in cosmological models. Can it lead to a regular
evolution, provided we manage to arrange it in some way? The answer is
no, for the following intuitive reason: We can get $p_{\phi}$ to be
very small by making special choices, but it will not be exactly zero
and eventually the diverging $a^{-3}$ will win if we only go close
enough to $a=0$. In the end, we always have to face the singularity
problem illustrated by the simple solution $a(t)\propto\sqrt{t-t_0}$
for radiation: $a(t_0)=0$ such that all of space collapses to a single
point (any length of a space-like curve at $t_0$ measured with the
line element (\ref{dsflat}) is zero) and the energy density
diverges. The most dooming consequence is that the evolution breaks
down: We cannot set up an initial value problem at $t_0$ and evolve to
values of $t$ smaller than $t_0$. The theory does not tell us what
happens beyond $t_0$. This consequence is a general property of
general relativity which cannot be avoided. We used the symmetric
situation only for purposes of illustration, but the singularity
problem remains true for any solution \cite{HawkingEllis}. There will
always be points which can be reached in a finite amount of time, but
we will not be able to know anything as to what happens beyond such a
point. General relativity cannot be complete since it predicts
situations where it breaks down.

This is the classical situation. Can it be better in a quantum theory
of gravity? In fact, this has been the hope for decades, justified by
the following motivation: The classical hydrogen atom is unstable, but
we know well that quantum mechanics leads to a ground state of finite
energy $E_0=-\frac{1}{2}m_{\rm e}e^4/\hbar^2$ which cures the
instability problem. One can easily see that this is the only
non-relativistic energy scale which can be built from the fundamental
parameters purely for dimensional reasons. In particular, a non-zero
$\hbar$ is necessary, for in the classical limit $\hbar\to 0$ the
ground state energy diverges leading to the classical instability. As
an additional consequence we know that the existence of a non-zero
$\hbar$ leads to discrete energies.

In gravity the situation is similar. We have its fundamental parameter
$G$ from which we can build a natural length scale\footnote{Sometimes the Planck length is defined as $\sqrt{G\hbar}$.} 
$\lP=\sqrt{8\pi G\hbar}$, the Planck length. It is very tiny and becomes important
only at small scales, e.g.\ close to classical singularities where the
whole space is small. Where the Planck length becomes important we
expect deviations from the classical behavior which will hopefully
cure the singularity problem. In the classical limit, $\hbar\to0$, the
Planck length becomes zero and we would get back the
singularity. Completing our suggestions from the hydrogen atom, we
also expect discrete lengths in a quantum theory of gravity, the
explicit form of which can only be concluded from a precise
implementation.

\section{Wheeler--DeWitt quantum gravity}
\label{s:WdW}

As discussed, one can bring general relativity into a canonical
formulation where the metric $q_{ab}$ and its momenta $\pi^{ab}$ play
the role of phase space coordinates (infinitely many because they
depend on the points of space), together with possible matter degrees
of freedom and their momenta. This allows us to perform a canonical
quantization (see, e.g., \cite{QCReview}) by representing quantum
states as functionals $\Psi(q_{ab},\phi)$ of the metric and matter
fields, corresponding to a metric representation. The metric itself
then acts as a multiplication operator, and its conjugate $\pi^{ab}$
by a functional derivative $\hat{\pi}^{ab}=-\I\hbar\partial/\partial
q_{ab}$. These are the basic operators from which more complicated
ones can be constructed.

\subsection{The Wheeler--DeWitt equation}

In a canonical formulation of general relativity, the dynamics is
determined by a constraint equation, (\ref{HADM}) in the variables
used here. Replacing $q_{ab}$ and $\pi^{ab}$ by the respective
operators yields a complicated constraint operator $\hat{H}_{\rm ADM}$
acting on a wave function $\Psi$. Since the classical expression must
vanish, only states $\Psi$ are allowed which are annihilated by the
constraint operator, i.e.\ they have to fulfill the Wheeler--DeWitt
equation $\hat{H}_{\rm ADM}\Psi=0$. Since the constraint is quadratic
in the momenta, this is a second order functional differential
equation. However, it is only formal since it contains products of
functional derivatives which have to be regularized in a way which
does not spoil the properties of the theory, in particular its
background independence. Such a regularization is complicated because
the classical constraint is not even a polynomial in the basic fields,
and so far it has not been done successfully in the ADM formulation.

There is another apparent difficulty with the constraint equation: It
is supposed to give us the dynamics, but there is no time dependence
at all, and no time derivative part as in a Schr\"odinger
equation. This is a general property of theories as general relativity
which are invariant under four-dimensional coordinate
transformations. We do not have an absolute notion of time, and thus
it cannot appear in the basic evolution equation. Classically, we can
introduce a time parameter (coordinate time $t$), but it just serves
to parameterize classical trajectories. It can be changed freely by a
coordinate transformation. In the quantum theory, which is formulated
in a coordinate independent way, coordinate time cannot appear
explicitly. Instead, one has to understand the evolution in a
relational way: there is no evolution with respect to an absolute
time, but only evolution of all the degrees of freedom with respect to
each other. After all, this is how we perceive time. We build a clock,
which is a collection of matter degrees of freedom with very special
interactions with each other, and observe the evolution of other
objects, degrees of freedom with weak interactions with the clock,
with respect to its progression. Similarly, we can imagine to select a
particular combination of matter or metric degrees of freedom as our
clock variable and re-express the constraint equation as an evolution
equation with respect to it \cite{DeWitt,Time}. For instance, in a cosmological context
we can choose the volume of space as {\em internal time\/} and measure
the evolution of matter degrees of freedom with respect to the
expansion or contraction of the universe. In general, however, a
global choice of a time degree of freedom which would allow us to
bring the full Wheeler--DeWitt equation into the form of an evolution
equation, is not known; this is the problem of time in general
relativity.

Due to the complicated regularization and interpretational issues,
applications of the full Wheeler--DeWitt equation have been done only
at a formal level for semiclassical calculations.

\subsection{Minisuperspaces}

In order to study the theory explicitly, we again have to resort to
approximations. A common simplification of the Wheeler--DeWitt
formalism is the reduction to minisuperspace models where the space is
homogeneous or even isotropic. Therefore, the metric of space is
specified by a finite number of parameters only -- only the scale
factor $a$ in the isotropic case. While this is similar in spirit to
looking for symmetric classical solutions as we did in section
\ref{s:GR}, there is also an important difference: If we want the
symmetry to be preserved in time we need to restrict the time
derivative of the metric, i.e.\ its canonical conjugate, in the same
symmetric form. This is possible classically, but in quantum mechanics
it violates Heisenberg's uncertainty relations for the excluded
degrees of freedom. Minisuperspace models do not just give us particular,
if very special exact solutions as in the classical theory; their
results must be regarded as approximations which are valid only under
the assumption that the interaction with the excluded parameters is
negligible.

An isotropic minisuperspace model has the two gravitational parameters
$a$ and its conjugate $p_a=3a\dot{a}/8\pi G$ together with possible
matter degrees of freedom which we simply denote as $\phi$ and
$p_{\phi}$. Using a Schr\"odinger quantization of the momenta acting
on a wave function $\psi(a,\phi)$, the Friedmann equation
(\ref{Friedmann}) is quantized to the Wheeler--DeWitt equation
\begin{equation} \label{WdW}
 \frac{3}{2}\left(-\frac{1}{9}\lP^4 a^{-1}\frac{\partial}{\partial a}a^{-1}
  \frac{\partial}{\partial a}+k\right) a \psi(a,\phi)= 8\pi G\hat{H}_{\phi}(a)
  \psi(a,\phi)
\end{equation}
with matter Hamiltonian $\hat{H}_{\phi}(a)$. This equation is not
unique due to ordering ambiguities on the left hand side. Here, we use
the one which is related to the quantization derived later. Without
fixing the ordering ambiguity, consequences derived from the equation
are ambiguous \cite{Konto}.

The Wheeler--DeWitt equation quantizes the dynamical classical
equation and thus should describe the quantum dynamics. 
As described before,
in an isotropic model we can select the scale factor $a$ as an internal
time; evolution of the matter fields will then be measured not in
absolute terms but in relation to the expansion or contraction of the
universe. Interpreting $a$ as a time variable immediately brings
Eq.~(\ref{WdW}) to the form of a time evolution equation, albeit with
an unconventional time derivative term.

An unresolvable problem of the Wheeler--DeWitt quantization, however,
is that it is still singular. Energy densities, all depending on the
multiplication operator $a^{-1}$ are still unbounded, and the
Wheeler--DeWitt equation does not tell us what happens at the other
side of the classical singularity at $a=0$. Instead, the point of view
has been that the universe is ``created'' at $a=0$ such that initial
conditions have to be imposed there. DeWitt \cite{DeWitt} tried to
combine both problems by requiring $\psi(0)=0$ which can be
interpreted as requiring a vanishing probability density to find the
universe at the singularity. However, this very probability
interpretation, which is just taken over from quantum mechanics, is
not known to make sense in a quantum cosmological
context. Furthermore, at the very least one would also need
appropriate fall-off conditions for the wave function since otherwise
we can still get arbitrarily close to the singularity. Appropriate
conditions are not known, and it is not at all clear if they could
always be implemented. The worst problem is, however, that DeWitt's
initial condition is not well-posed in more general models where its
only solution would vanish identically.

DeWitt's condition has been replaced by several proposals which are
motivated from different intuitions
\cite{tunneling,nobound}. However, they do not eliminate the
singularity; with them the wave function would not even vanish at
$a=0$ in the isotropic case. They accept the singularity as a point of
creation.

Thus, the hope motivated from the hydrogen atom has not
materialized. The isotropic universe model is still singular after
quantizing. Do we have to accept that singularities of gravitational
systems cannot be removed, not even by quantization? If the answer would
be affirmative, it would spell severe problems for any desire to
describe the real world by a physical theory. It would mean that there
can be no complete description at all; any attempt would stop at the
singularity problem.

Fortunately, the answer turns out not to be affirmative. We will see
in the following sections that singularities are removed by an
appropriate quantization. Why, then, is this not the case for the
Wheeler--DeWitt quantization? One has to keep in mind that there is no
mathematically well-defined Wheeler--DeWitt quantization of full
general relativity, and systematic investigations of even the formal
equations are lacking. What one usually does instead is merely a
quantum mechanical application in a simple model with only a few
degrees of freedom. There is no full theory which one could use to see
if all quantization steps would also be possible there. General
relativity is a complicated theory and its quantization can be done,
if at all, only in very special ways which have to respect complicated
consistency conditions, e.g.\ in the form of commutation relations
between basic operators. In a simple model, all these problems can be
brushed over and consistency conditions are easily overlooked. One
hint that this in fact happened in the Wheeler--DeWitt quantization is
the lacking discreteness of space. We expected that a non-zero Planck
length in quantum gravity would lead to the discreteness of
space. While we did see the Planck length in Eq.~(\ref{WdW}), there
was no associated discreteness: the scale factor operator, which is
simply the multiplication operator $a$, still has continuous spectrum.

After the discussion it should now be clear how one has to proceed in
a more reliable way. We have to use as much of the full theory of
quantum gravity as we know and be very careful to use only techniques
in our symmetric models which can also be implemented in the full
theory. In this way, we would respect all consistency conditions and
obtain a faithful model of the full theory. Ideally, we would even
start from the full theory and define symmetric models there at the
level of states and operators.

By now, we have good candidates for a full theory of quantum gravity,
and in the case of quantum geometry \cite{Nonpert,Rov:Loops,ThomasRev}
also a procedure to define symmetric models from it \cite{SymmRed}. We
will describe the main results in the following two sections.

\section{Quantum geometry}
\label{s:QG}

While the Wheeler--DeWitt quantization is based on the ADM formulation
whose basic variables are the metric and extrinsic curvature of a
spatial slice, quantum geometry is a newer approach based on
Ashtekar's variables. Quantization, in particular of a non-linear
field theory, is a delicate step whose success can depend
significantly on the formulation of a given classical
theory. Classical theories can usually be formulated in many
different, equivalent ways, all being related by canonical
transformations. Not all of these transformations, however, can be
implemented unitarily at the quantum level which would be necessary
for the quantum theories to be equivalent, too. For instance, when
quantizing one has to choose a set of basic variables closed under
taking Poisson brackets which are promoted unambiguously to operators
in such a way that their Poisson brackets are mapped to commutation
relations. There is no unique prescription to quantize other functions
on phase space which are not just linear functions of the basic ones,
giving rise to quantization ambiguities. In quantum mechanics one can
give quite general conditions for a representation of at least the
basic variables to be unique (this representation is the well-known
Schr\"odinger quantization). However, such a theorem is not available
for a field theory with infinitely many degrees of freedom such that
even the basic variables cannot be quantized uniquely without further
conditions. 

One can often use symmetry requirements together with other natural conditions in order to select a unique
representation of the basic variables, e.g.\ Poincar\'e invariance for
a field theory on Minkowski space as a background \cite{Haag}. For general
relativity, which is background independent, it has recently been
proven in the context of quantum geometry that diffeomorphism
invariance, i.e.\ invariance under arbitrary deformations of space,
can replace Poincar\'e invariance in strongly restricting the class of possible representations
\cite{Rep}. It is clear that those
precise theorems can only be achieved within a theory which is
mathematically well-defined. The Wheeler--DeWitt quantization, on the
other hand, does not exist beyond a purely formal level and it is
unknown if it can give a well-defined quantum representation of the
ADM variables at all. In any case, it is based on basic variables
different from the ones quantum geometry is based on so that any
representation it defines would likely be inequivalent to the one of
quantum geometry.

From the beginning, quantum geometry was striving for a mathematically
rigorous formulation. This has been possible because it uses
Ashtekar's variables which bring general relativity into the form of a
gauge theory. While not all standard techniques for quantizing a gauge
theory can be applied (most of them are not background independent),
new powerful techniques for a {\em background independent\/}
quantization have been developed \cite{ALMMT,DiffGeom,FuncInt}. This
was possible only because the space of connections, which is the
configuration space of quantum geometry, has a structure much better
understood than the configuration space of the Wheeler--DeWitt
quantization, namely the space of metrics.

We do not describe those techniques here and instead refer the
interested reader to the literature where by now several technical
reviews are available \cite{ThomasRev,ALRev}. In this section, instead, we
present an intuitive construction which illustrates all the main
results.

\subsection{Basic operators and states}

As usually in gauge theories (for instance in lattice formulations),
one can form holonomies as functions of connections for all curves
$e\colon [0,1]\to\Sigma$
in a manifold $\Sigma$,
\begin{equation}
 h_e(A)={\cal P}\exp\left(\int_eA_a^i(e(t))\,\dot{e}^a(t)\tau_i\D
 t\right) \in{\rm SU(2)}
\end{equation}
where $\dot{e}^a$ is the tangent vector to the curve $e$ and
$\tau_i=-\frac{\I}{2}\sigma_i$ are generators of the gauge group SU(2)
in terms of the Pauli matrices. The symbol ${\cal P}$ denotes path
ordering which means that the non-commuting su(2) elements in the
exponential are ordered along the curve. Similarly, given a surface
$S\colon [0,1]^2\to\Sigma$ we can form a flux as a function of the
triads,
\begin{equation}
 E(S)=\int_SE^a_i(y)n_a(y)\tau^i\D^2y
\end{equation}
where $n_a$ is the co-normal\footnote{The co-normal is defined as
$n_a=\frac{1}{2}\epsilon_{abc}\epsilon^{de}(\partial x^b/\partial
y^d)(\partial x^c/\partial y^e)$ without using a background metric,
where $x^a$ are coordinates of $\Sigma$ and $y^d$ coordinates of the
surface $S$.} to the surface $S$. Holonomies and fluxes are the basic
variables which are used for quantum geometry, and they represent the
phase space of general relativity faithfully in the sense that any two
configurations of general relativity can be distinguished by
evaluating holonomies and fluxes in them.

One can now prove that the set of holonomies and fluxes is closed
under taking Poisson brackets and that there is a representation of
this Poisson algebra as an operator algebra on a function
space. Moreover, using the action of the diffeomorphism group on
$\Sigma$, which deforms the edges and surfaces involved in the above
definitions, this representation is the unique covariant one. Note
that unlike the functional derivatives appearing in the Wheeler-DeWitt
quantization, these are well-defined operators on an infinite
dimensional Hilbert space. Note in particular that holonomies are
well-defined as operators, but {\em not\/} the connection itself. A
Wheeler--DeWitt quantization, on the other hand, regards the extrinsic
curvature, related to the connection, as one of the basic fields and
would try to promote it to an operator. This is not possible in
quantum geometry (and it is not known if it is possible at a precise
level at all) which demonstrates the inequivalence of both
approaches. The fact that only the holonomies can be quantized can
also be seen as one of the consistency conditions of a full theory of
quantum gravity mentioned earlier. In a minisuperspace model one can
easily quantize the isotropic extrinsic curvature, which is
proportional to $p_a/a$. However, since it is not possible in the full
theory, the model departs from it already at a very basic level. A
reliable model of a quantum theory of gravity should implement the
feature that only holonomies can be quantized; we will come back to
this issue later.

We did not yet specify the space of functions on which the basic
operators act in the representation of quantum
geometry. Understandably, a full definition involves many techniques
of functional analysis, but it can also be described in intuitive
terms. As mentioned already, it is convenient to define the theory in
a connection representation since the space of connections is
well-understood. We can then start with the function ${\mathbf 1}$
which takes the value one in every connection and regard it as our
ground state.\footnote{Note that we do not call it ``vacuum state''
since the usual term ``vacuum'' denotes a state in which matter is
unexcited but the gravitational background is Minkowski space (or
another non-degenerate solution of general relativity). We will see
shortly, however, that the ground state we are using here represents a
state in which even gravity is ``unexcited'' in the sense that it
defines a completely degenerate geometry.} The holonomies depend only
on the connection and thus act as multiplication operators in a
connection formulation \cite{LoopRep}. Acting with a single holonomy
$h_e(A)$ on the state ${\mathbf 1}$ results in a state which depends
on the connection in a non-trivial way, but only on its values along
the curve $e$.  More precisely, since holonomies take values in the
group SU(2), we should choose an SU(2)-representation, for instance
the fundamental one, and regard the matrix elements of the holonomy in
this representation as multiplication operators. This can be done with
holonomies along all possible curves, and also acting with the same
curve several times. Those operators can be regarded as basic creation
operators of the quantum theory. Acting with holonomies along
different curves results in a dependence on the connection along all
the curves, while acting with holonomies along the same curve leads to
a dependence along the curve in a more complicated way given by
multiplying all the fundamental representations to higher ones. One
can imagine that the state space obtained in this way with all
possible edges (possibly intersecting and overlapping) in arbitrary
numbers is quite complicated, but not all states obtained in this way
are independent: one has to respect the decomposition rules of
representations. This can all be done resulting in a basis of states,
the so-called spin network states \cite{RS:Spinnet}. Furthermore, they
are orthonormal with respect to the diffeomorphism invariant measure
singled out by the representation, the Ashtekar--Lewandowski measure
\cite{DiffGeom}.

Note also that the quantum theory should be invariant under
SU(2)-rotations of the fields since a rotated triad does not give us a
new metric. Holonomies are not gauge invariant in this sense,
but as in lattice gauge theories we can use Wilson loops instead which
are defined as traces of holonomies along a closed loop. Repeating the
above construction only using Wilson loops results in gauge invariant
states.

Since we used holonomies to construct our state space, their action
can be obtained by multiplication and subsequent decomposition in the
independent states. Fluxes, on the other hand, are built from the
conjugate of connections and thus become derivative operators. Their
action is most easy to understand for a flux with a surface $S$ which
is transversal to all curves used in constructing a given state. Since
the value of a triad in a given point is conjugate to the connection
in the same point but Poisson commutes with values of the connection
in any other point, the flux operator will only notice {\em
intersection points\/} of the surface with all the edges which will be
summed over with individual contributions.  The contributions of all
the intersection points are the same if we count intersections with
overlapping curves separately. In this way, acting with a flux
operator on a state returns the same state multiplied with the
intersection number between the surface of the flux and all the curves
in the state. This immediately shows us the eigenvalues of flux
operators which turn out to be {\em discrete}. Since the fluxes are
the basic operators representing the triad from which geometric
quantities like length, area and volume are constructed, it shows that
geometry is discrete \cite{AreaVol,Area,Vol2,Len}. The main part of the
area spectrum for a given surface $S$ (the one disregarding
intersections of curves in the state) is
\begin{equation}
 A(S)= \gamma\lP^2\sum_i \sqrt{j_i(j_i+1)}
\end{equation}
where the sum is over all intersections of the surface $S$ with curves
in the state, and the SU(2)-labels $j_i$ parameterize the multiplicity
if curves overlap (without overlapping curves, all $j_i$ are
$\frac{1}{2}$). Thus, quantum geometry predicts that geometric spectra
are discrete, and it also provides an explicit form. Note that the
Planck length appears (which arises because the basic Poisson brackets
(\ref{Poisson}) contain the gravitational constant while $\hbar$
enters by quantizing a derivative operator), but the scale of the
discreteness is set by the Barbero--Immirzi parameter $\gamma$. While
different $\gamma$ lead to equivalent classical theories, the value of
the parameter does matter in the quantum theory. If $\gamma$ would be
large the discreteness would be important already at large scales
despite the smallness of the Planck length. Calculations from black
hole entropy, however, show that $\gamma$ must be smaller than one,
its precise value being $\log(2)/\pi\sqrt{3}$
\cite{Entro}.

 Thus, quantum geometry already fulfills one of our expectations of
 Section \ref{s:GR}, namely that quantum gravity should predict a
 discreteness of geometry with a scale set roughly by the Planck
 length. Note that the use of holonomies in constructing the quantum
 theory, which was necessary for a well-defined formulation, is
 essential in obtaining the result about the discreteness. This fact
 has been overlooked in the Wheeler--DeWitt quantization which,
 consequently, does not show the discreteness.

\subsection{Composite operators}

We now have a well-defined framework with a quantization of our basic
quantities, holonomies and fluxes. Using them we can construct
composite operators, e.g.\ geometric ones like area and volume or the
constraint operator which governs the dynamics. Many of them have
already been defined,
but they are quite complicated. The volume operator, for instance, has
been constructed and it has been shown to have a discrete spectrum
\cite{AreaVol,Vol2,Vol}; determining all its eigenvalues, however, would
require the diagonalization of arbitrarily large matrices. Since it
plays an important role in constructing other operators, in particular
the Hamiltonian constraint \cite{QSDI,QSDV}, it makes explicit
calculations in the whole theory complicated.

The constraint, for instance can be quantized by using a small Wilson
loop along some loop $\alpha$, which has the expansion
$h_{\alpha}=1+As_1^as_2^bF_{ab}^i\tau_i$ where $A$ is the coordinate
area of the loop and $s_1$ and $s_2$ are tangent vectors to two of its
edges, to quantize the curvature of the connection. The product of
triads divided by the determinant appears to be problematic because
the triad can be degenerate resulting in a vanishing
determinant. However, one can make use of the classical identity
\cite{QSDI}
\begin{equation} \label{iden}
 \frac{E^a_iE^b_j\epsilon^{ijk}}{\sqrt{|\det E|}}=
 \frac{1}{4\pi \gamma G}\epsilon^{abc} \{A_c^k,V\}\;,
\end{equation}
replace the connection components by holonomies $h_s$ and use the volume
operator to quantize this expression in a non-degenerate way. For the first part\footnote{The remaining part of the constraint involving extrinsic curvature components can be obtained from the first part since the extrinsic curvature can be written as a Poisson bracket of the first part of the constraint with the volume \cite{QSDI}.}  of the constraint (\ref{HAsh}) this results in
\begin{equation}
 \hat{H}=\sum_{v\in {\cal V}} \sum_{v(\Delta)=v} \epsilon^{IJK} \tr(h_{\alpha_{IJ}(\Delta)}h_{s_K(\Delta)} 
 [h_{s_K(\Delta)}^{-1},\hat{V}_v])
\end{equation}
where we sum over the set ${\cal V}$ of vertices of the graph belonging to the state we act on, and over all possible choices (up to diffeomorphisms) to form a tetrahedron $\Delta$ with a loop $\alpha_{IJ}(\Delta)$ sharing two sides with the graph and a third transversal curve $s_K(\Delta)$. The first holonomy along $\alpha_{IJ}(\Delta)$ quantizes the curvature components while $h_{s_K(\Delta)}$ together with the commutator quantizes the triad components.

A similar strategy can be used for matter Hamiltonians which usually
also require to divide by the determinant of the triad at least in
their kinetic terms. Here it is enough to cite some operators as examples to illustrate the general structure which will later be used in Section \ref{s:Phen}; for details we refer to \cite{QSDV}. For electromagnetism, for instance, we need to quantize
\begin{eqnarray}  \label{hmax}
 H_{{\rm Maxwell}} &=& \int_{\Sigma} \D^3x \;  \frac{q_{ab}}{2Q^2\sqrt{\det q}}
 [\underline E ^a \underline E^b + \underline B ^a \underline B^b]
\end{eqnarray}
where $(\underline{A}_a, \underline E^a/Q^2)$ are the canonical fields of the electromagnetic sector with gauge group U(1) and coupling
constant $Q$, related to the dimensionless fine structure constant by 
$\alpha_{\rm EM}=Q^2\hbar$. Furthermore, $\underline B^b$ is the magnetic field of
the U(1) connection $\underline A$, i.e.\ the dual of its curvature.

Along the lines followed for the gravitational Hamiltonian we obtain the full electromagnetic Hamiltonian operator \cite{QSDV} (with a weight factor $w(v)$ depending on the graph)
\begin{eqnarray}\label{HamEM}
\hat{H}_{\rm Maxwell} &=&\frac{1}{2\lP^{4}Q^{2}}\sum_{v\in {\cal V}}\,w(v)\sum_{v(\Delta )=v({%
\Delta ^{\prime }})=v}  
\tr\left( \tau _{i}\,h_{s_{L}(\Delta )}\left[ h_{s_{L}(\Delta
)}^{-1},\sqrt{{\hat{V}}_{v}}\right] \right)\nonumber   \\
&&\times \tr\left( \tau
_{i}h_{s_{P}(\Delta ^{\prime })}\left[ h_{s_{P}(\Delta ^{\prime })}^{-1},%
\sqrt{{\hat{V}}_{v}}\right] \right) \,  \nonumber \\
&&\times \epsilon ^{JKL}\epsilon ^{MNP}\left[ \left( \E^{-\I{\hat{\Phi}}%
^{B}_{JK}}-1\right) \left( \E^{-\I{\hat{\Phi}}^{\prime B}_{MN}
}-1\right) -\hat{\Phi}^{E}_{JK}\hat{\Phi}^{\prime E}_{MN})%
\right].\nonumber \\ \label{EHC}
\end{eqnarray}
Let us emphasize the structure of the above regularized Hamiltonian. There is a common
gravitational factor included in the SU(2) trace. The basic entities that
quantize the electromagnetic part are the corresponding fluxes (as operators acting on a state for the electromagnetic field): one is associated
with the magnetic field, which enters through a product of exponential flux factors $\exp(-\I\Phi^{(\prime)B})$ constructed from holonomies in $\Delta$ and $\Delta'$, respectively, while
the other is related to the electric field, entering in a bilinear product of electric fluxes $\Phi^{(\prime)E}$. 

Similarly, one can set up a theory for fermions which would be coupled to the gauge fields and to gravity. For a spin-$\frac{1}{2}$ field $\theta_A$ one obtains the kinetic part (with the Planck mass $m_{\rm P}=\hbar/\lP$)
\begin{eqnarray}\label{HamSpin}
\hat{H}_{\rm spin-1/2} &=& - \frac{m_{\rm P}}{2\lP^3} \sum_{v\in {\cal V}}  \sum_{v(\Delta)=v}
\epsilon^{ijk} \epsilon^{IJK} 
\tr \left( \tau_i h_{s_I(\Delta)}\left[h^{-1}_{s_I(\Delta)}, \sqrt{\hat V_v)}\right]\right) \nonumber\\
&&\times \tr \left( \tau_j h_{s_J(\Delta)}\left[h^{-1}_{s_J(\Delta)}, \sqrt{\hat V_v)}\right]\right)\nonumber \\
&&  \times \left[\left[(\tau_k h_{s_K(\Delta)}\theta)|_{s_K(\Delta)}-\theta|_v\right]_A\frac{\partial}{\partial\theta_A(v)} + h.c.\right]\;.
\end{eqnarray}

All of these matter Hamiltonians are
well-defined, bounded operators, which is remarkable since in quantum
field theories on a classical background matter Hamiltonians usually have ultraviolet
divergences. This can be interpreted as a natural cut-off implied by
the discrete structure. Compared to the Wheeler--DeWitt quantization it is a huge progress
that well-defined Hamiltonian constraint operators are available in
the full theory. Not surprisingly, their action is very complicated
for several reasons. The most obvious ones are the fact that Wilson
loops necessary to quantize curvature components create many new
curves in a state which is acted on, and that the volume operator is
being used to quantize triad components. The first fact implies that
complicated graphs are created, while the second one shows that even a
single one of those contributions is difficult to analyze due to the
unknown volume spectrum. And after determining the action of the
constraint operator on states we still have to solve it, i.e.\ find
its kernel. Furthermore, there are always several possible ways to quantize a classical Hamiltonian such that the ones we wrote down should be considered as possible choices which incorporate the main features.

The complicated nature should not come as a
surprise, though. After all, we are dealing with a quantization of full
general relativity without any simplifying assumptions. Even the
classical equations are difficult to solve and to analyze if we do not
assume symmetries or employ approximation schemes.
Those simplifications are also available for quantum geometry, which
is the subject of the rest of this article. Symmetries can be
introduced at the level of states which can be rigorously defined as
distributional, i.e.\ non-normalizable states (they cannot be ordinary
states since the discrete structure would break any continuous
symmetry). Approximations can be done in many ways, and different
schemes are currently being worked out. 

\section{Loop quantum cosmology}
\label{s:Cosmo}

Loop quantum cosmology aims to investigate quantum geometry in
simplified situations which are obtained by implementing
symmetries. In contrast to a Wheeler--DeWitt quantization and its
minisuperspace models, there is now also a full theory available. It
is then possible to perform all the steps of the quantization in a
manner analogous to those in the full theory. In particular, one can
be careful enough to respect all consistency conditions as, e.g., the
use of holonomies.\footnote{This sometimes requires to perform
manipulations which seem more complicated than necessary or even
unnatural from the point of view of a reduced model. However, all of
this can be motivated from the full theory, and in fact exploiting simplifications
which are not available in a full theory can always be
misleading.} There is a tighter relation between symmetric models and
the full theory which goes beyond pure analogy. For instance,
symmetric states can be defined rigorously \cite{SymmRed,PhD,cosmoI} and
the relation between operators is currently being investigated. In
this section, as already in the previous one, we use intuitive ideas
to describe the results.

In addition to testing implications of the full theory in a simpler
context, it is also possible to derive physical results. Fortunately,
many interesting and realistic physical situations can be approximated
by symmetric ones. This is true in particular for cosmology where one
can assume the universe to be homogeneous and isotropic at large scales.

\subsection{Symmetric states and basic operators}

As seen before, the canonical fields of a theory of gravity are completely
described by two numbers (depending on time) in an isotropic
context. For Ashtekar's variables, isotropic connections and triads
take the form
\begin{equation} \label{Inv}
 A_a^i(x)\D x^a=c\, \omega^i \quad,\quad E^a_i\frac{\partial}{\partial x^a}
 = p\, X_i
\end{equation}
where $\omega^i$ are invariant 1-forms and $X_i$ invariant vector
fields. For a spatially flat configuration, $\omega^i=\D x^i$ are just
coordinate differentials, while $X_i$ are the derivatives; for
non-zero spatial curvature the coordinate dependence is more complicated.
In all isotropic models, $c$ and $p$ are functions just of time. Their
relation to the isotropic variables used before is
\begin{equation}
 c={\textstyle\frac{1}{2}} (k-\gamma\dot{a}) \quad,\quad |p|=a^2
\end{equation}
such that $\{c,p\}=(8\pi/3)\gamma G$.  An important difference
is that $p$ can have both signs, corresponding to the two possible
orientations of a triad. In a classical theory we would ultimately
have to restrict to one sign since $p=0$ represents a degenerate
triad, and positive and negative signs are disconnected. But the
situation can (and will) be different in a quantum theory.

We can now perform an analog of the construction of states in the full
theory. The symmetry condition can be implemented by using only
invariant connections (\ref{Inv}) in holonomies as creation operators,
i.e.\
\[
 h_i(c)=\exp(c\tau_i)= \cos(c/2)+2\tau_i\sin(c/2)\;.
\]
Consequently, all the states we construct by acting on the
ground state ${\mathbf 1}$ will be functions of only the variable
$c$. All the complication of the full states with an arbitrary number
of curves has collapsed because of our symmetry assumption. The analog
of the spin network basis, an orthonormal basis in the connection
representation, is given by\footnote{A more careful analysis shows
that the Hilbert space of loop quantum cosmology is not separable
\cite{Bohr}. For our purposes, however, we can restrict to the
separable subspace used here which is left invariant by our
operators.} \cite{IsoCosmo}
\begin{equation}
 \langle c|n\rangle = \frac{\exp(\I nc/2)}{\sqrt{2}\sin(c/2)}
\end{equation}
for all integer $n$.

An analog of the flux operator is given by a quantization of the
isotropic triad component $p$,
\begin{equation}\label{p}
 \hat{p}|n\rangle = {\textstyle\frac{1}{6}}\gamma\lP^2 n|n\rangle\;.
\end{equation}
This immediately allows a number of observations: Its spectrum is
discrete ($n$ corresponds to the intersection number in the full
theory) which results in a discrete geometry. The scale factor
$\hat{a}=\sqrt{|\hat{p}|}$ also has a discrete spectrum which is very
different from the Wheeler--DeWitt quantization where the scale factor
is just a multiplication operator with a continuous spectrum. Thus, we
obtain a different quantization with deviations being most important
for small scale factors, close to the classical singularity. The
quantization $\hat{p}$ also tells us that the sign of the
``intersection number'' $n$ determines the orientation of space. There
is only one state, $|0\rangle$, which is annihilated by $\hat{p}$; we
identify it with the classical singularity.

We can also use $\hat{p}$ in order to obtain a quantization of the
volume \cite{cosmoII}; here we use the convention
\begin{equation}\label{Vol}
 V_{(|n|-1)/2}=\left({\textstyle\frac{1}{6}}\gamma\lP^2\right)^{\frac{3}{2}}
 \sqrt{(|n|-1)|n|(|n|+1)}
\end{equation}
for its eigenvalue on a state $|n\rangle$. Thus, we achieved our aims:
We have a quantization of a model which simplifies the full
theory. This can be seen from the simple nature of the states and the
explicit form of the volume spectrum which is not available in the
full theory. As we will see later, this also allows us to derive
explicit composite operators \cite{cosmoIII,IsoCosmo}. At the same
time, we managed to preserve essential features of the full theory
leading to a quantization different from\footnote{Both quantizations
are in fact inequivalent because, e.g., the operator $\hat{a}$ has
discrete spectrum in one and a continuous spectrum in the other
quantization.} the Wheeler--DeWitt one which lacks a relation to a
full theory.

\subsection{Inverse powers of the scale factor}

We can now use the framework to perform further tests of essential
aspects of quantum cosmology. In the Wheeler--DeWitt quantization we
have seen that the inverse scale factor $a^{-1}$ becomes an unbounded
operator. Since its powers also appear in matter Hamiltonians, their
quantizations will also be unbounded, reflecting the classical
divergence. The situation in loop quantum cosmology looks even worse
at first glance: $\hat{a}$ still contains zero in its spectrum, but
now as a discrete point. Then its inverse does not even exist. There
can now be two possibilities: It can be that we cannot get a
quantization of the classically diverging $a^{-1}$, which would mean
that there is no way to resolve the classical singularity. As the other
possibility it can turn out that there are admissible quantizations of
$a^{-1}$ in the sense that they have the correct classical limit and
are densely defined operators. The second possibility exists because,
as noted earlier, there are usually several possibilities to construct
a non-basic operator like $a^{-1}$. If the simplest one fails (looking
for an inverse of $\hat{a}$), it does not mean that there is no
quantization at all.

It turns out that the second possibility is realized \cite{InvScale}, in a
way special to quantum geometry. We can use the identity (\ref{iden}),
which has been essential in quantizing Hamiltonians, in order to
rewrite $a^{-1}$ in a classically equivalent form which allows a
well-defined quantization. In this way, again, we stay very close to
the full theory, repeating only what can be done there, and at the
same time obtain physically interesting results.

The reformulation can be written in a simple way for a symmetric
context, e.g.,\footnote{One can easily see that there are many ways
to rewrite $a^{-1}$ in such a way. Essential features, however, are
common to all these reformulations, for instance the fact that always
the absolute value of $p$ will appear in the Poisson bracket rather
than $p$ itself. We will come back to this issue shortly in the
context of quantization ambiguities.}
\begin{eqnarray}\label{reform}
 a^{-1} &=& \left(\frac{1}{2\pi\gamma G} \{c,|p|^{3/4}\}\right)^2=
 \left(\frac{1}{3\pi\gamma G}\sum_i\tr(\tau_i h_i(c)
 \{h_i(c)^{-1},\sqrt{V}\})\right)^2\nonumber\\
 &=&  \left((2\pi\gamma G\; j(j+1)(2j+1))^{-1} \sum_i\tr_j(\tau_i h_i(c)
 \{h_i(c)^{-1},\sqrt{V}\})\right)^2
\end{eqnarray}
indicating in the last step that we can choose any
SU(2)-representation when computing the trace without changing the
classical expression (the trace without label is in the fundamental
representation, $j=\frac{1}{2}$).  Note that we only need a positive
power of $p$ at the right hand side which can easily be quantized. We
just have to use holonomy operators and the volume operator, and
turn the Poisson bracket into a commutator. This results in a
well-defined operator which has eigenstates $|n\rangle$ and, for
$j=\frac{1}{2}$, eigenvalues
\begin{equation} \label{ainv}
 \widehat{a^{-1}}|n\rangle = \frac{16}{\gamma^2\lP^4} \left(
 \sqrt{V_{|n|/2}}- \sqrt{V_{|n|/2-1}}\right)^2 |n\rangle
\end{equation}
in terms of the volume eigenvalues (\ref{Vol}).

It has the following properties \cite{InvScale,Ambig}; see Fig.~\ref{Ainv}:

\begin{figure}
\centering
 \includegraphics[height=6cm]{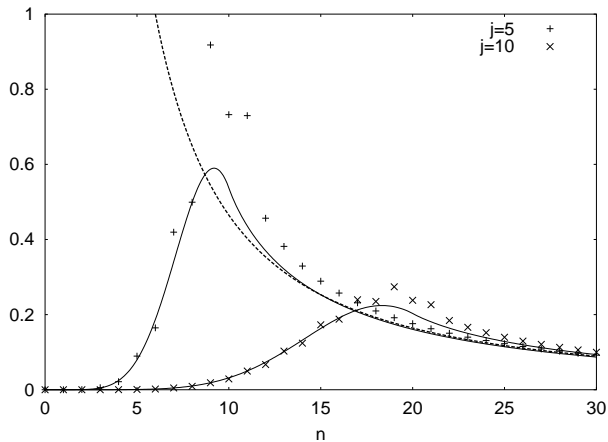}
\caption{Two examples for eigenvalues of inverse scale factor
operators with $j=5$ ($+$) and $j=10$ ($\times$), compared to the
classical behavior (dashed) and the approximations (\ref{ainvm})
\cite{Ambig}.}
\label{Ainv}
\end{figure}

\begin{enumerate}
 \item It is a {\em finite\/} operator with upper bound\footnote{We
 only use the general form of the upper bound. Its precise value
 depends on quantization ambiguities and is not important in this
 context.} $(a^{-1})_{\rm max}= \frac{32(2-\sqrt{2})}{3\lP}$ at a peak
 at $n=2$. Now we can see that the situation is just as in the case of
 the hydrogen atom: The classically pathological behavior is cured by
 quantum effects which -- purely for dimensional reasons -- require
 $\hbar$ in the denominator (recall $\lP=\sqrt{8\pi G\hbar}$). A
 finite value for the upper bound is possible only with non-zero
 $\lP$; in the classical limit $\lP\to0$ we reobtain the classical
 divergence.

 \item The first point demonstrates that the classical behavior is
 modified at small volume, but one can see that it is approached
 rapidly for volumes larger than the peak position. Thus, the
 quantization (\ref{ainv}) has the correct classical limit and is
 perfectly admissible.

 \item While the first two points verify our optimistic expectations,
 there is also an unexpected feature. The classical divergence is not
 just cut off at a finite value, the eigenvalues of the inverse scale
 factor drop off when we go to smaller volume and are exactly zero for
 $n=0$ (where the eigenvalue of the scale factor is also zero). This
 feature, which will be important later, is explained by the fact that
 the right hand side of (\ref{reform}) also includes a factor of
 $\sgn(p)^2$ since the absolute value of $p$ appears in the Poisson
 bracket. Strictly speaking, we can only quantize $\sgn(p)^2a^{-1}$,
 not just $a^{-1}$ itself. Classically, we cannot distinguish
 between both expressions -- both are equally ill-defined for $a=0$
 and we would have to restrict to positive $p$. As it turns out,
 however, the expression with the sign does have well-defined
 quantizations, while the other one does not. Therefore, we have to
 use the sign when quantizing expressions involving inverse powers of
 $a$, and it is responsible for pushing the eigenvalue of the inverse
 scale factor at $n=0$ to zero.

 \item As already indicated in (\ref{reform}), we can rewrite the
 classical expression in many equivalent ways. Quantizations, however,
 will not necessarily be the same. In particular, using a higher
 representation $j\not=\frac{1}{2}$ in (\ref{reform}), the holonomies
 in a quantization will change $n$ by amounts larger than one. In
 (\ref{ainv}) we will then have volume eigenvalues not just with $n-1$
 and $n+1$, but from $n-2j$ to $n+2j$ corresponding to the coupling
 rules of angular momentum. Quantitative features depend on the
 particular value of $j$ (or other quantization ambiguities), but
 qualitative aspects -- in particular the ones in points 1 to 3 --
 {\em do not change}. Thus, the quantization is robust under
 ambiguities, but there can be small changes depending on which
 particular quantization is used. Such a freedom can also be exploited
 in a phenomenological analysis of some effects.
\end{enumerate}

Let us make the last point more explicit. The exact formula for
eigenvalues with a non-fundamental representation is quite
complicated. It can, however, be approximated using a rather simple
function \cite{Ambig}

\begin{figure}
\centering
 \includegraphics[height=6cm]{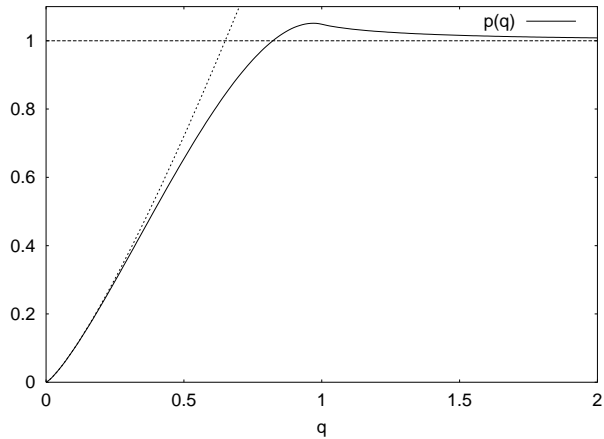}
\caption{The function $p(q)$ of (\ref{pq}), derived in \cite{Ambig}.
For small $q$, $p(q)$ increases like
$p(q)=\frac{12}{7}q^{5/4}(1-q+O(q^2))$ (dashed).}
\label{Prof}
\end{figure}

\begin{eqnarray} \label{pq}
 p(q) &=& {\textstyle\frac{8}{77}}\, q^{1/4}\left[ 7\left(
 (q+1)^{11/4}-|q-1|^{11/4}\right)\right.\nonumber\\
 && - \left.11q \left( (q+1)^{7/4} -
 \sgn(q-1) |q-1|^{7/4}\right)\right]\;,
\end{eqnarray}
see Fig.~\ref{Prof},
such that the eigenvalues of a quantization of $a^{-m}$ with positive
$m$ are given by
\begin{equation} \label{ainvm}
 (a^{-m})^{(j)}_n=V_{|n|/2}^{-m/3} p(|n|/2j)^{2m}
\end{equation}
with the ambiguity parameter $j$. There are many other ambiguities
which can change also the function $p$, but the one indicated by $j$
is most important. It parameterizes the position of the peak in an
inverse power of the scale factor, which roughly coincides with the
boundary between classical behavior and quantum modifications.

Note also that (\ref{ainvm}) displays the observation that inverse
powers of the scale factor annihilate the singular state $|0\rangle$,
thanks to $\lim_{q\to0}(q^{-1/4}p(q))=0$. This has important
consequences for matter Hamiltonians: they usually consist of a
kinetic term containing components of the inverse metric and a
potential term containing metric components. In the isotropic case, a
widely used example in cosmology is the scalar Hamiltonian
(\ref{Hphi}). We have already discussed the divergence at the
singularity of the classical kinetic term, unless $p_{\phi}=0$. The
potential term, on the other hand, vanishes there. When we quantize
this expression, we have to use an inverse power (\ref{ainvm}) in the
kinetic term; for the potential term we can just use volume
eigenvalues. Now, the potential term still vanishes at $n=0$, but so
does the kinetic term after quantization. Similarly, one can check
that any matter Hamiltonian $\hat{H}_{\rm matter}$ (assuming for
simplicity the absence of curvature couplings) fulfills
\begin{equation}
 \hat{H}_{\rm matter}|0\rangle=0
\end{equation}
irrespective of the particular kind of matter and its quantization.

All these observations indicate that the quantum behavior is much
better, less singular, than the classical one. The real test, however,
can only come from studying the quantum evolution. An absence of
singularities can be confirmed only if it is possible to extend the
evolution through the singular boundary; the theory has to tell us
what happens at the singularity and beyond.

\subsection{Dynamics\protect\footnote{A brief summary of the results in this subsection can be found in \cite{Essay}.}}

To study the dynamics of a theory we need its evolution equation which
for gravity is given by the Hamiltonian constraint. In the
Wheeler--DeWitt quantization we have seen that the constraint equation
takes the form of an evolution equation after quantizing in a metric
or triad representation and choosing an internal time $a$.

We can follow the same steps here if we first transform from the
connection representation used so far in quantum geometry to a triad
representation. This can be done straightforwardly since we already
know the triad eigenstates $|n\rangle$. A state $|\psi\rangle$ can
then be expanded in these eigenstates,
$|\psi\rangle=\sum_n\psi_n(\phi)|n\rangle$ denoting possible matter
degrees of freedom collectively by $\phi$. The coefficients
$\psi_n(\phi)$ in the expansion then define, as usually, the state in
the triad representation. Since $n$ denotes the eigenvalues of
$\hat{p}$, it will now play the role of an internal time. Here we
observe another difference to the Wheeler--DeWitt quantization: due to
the discrete geometry, also time is discrete in an internal time
picture.

The Wheeler--DeWitt quantization now proceeded by quantizing the
gravitational momentum $p_a$ by a differential operator as in quantum
mechanics. An analogous step is not possible in quantum geometry;
momenta here, i.e.\ connection components, have to be quantized using
holonomies which do not act as differential operators. Instead, they
act according to the SU(2) coupling rules, e.g.\
\begin{eqnarray*}
 \langle c|h_i(c)|n\rangle &=& \langle
 c|\cos(c/2)+2\tau_i\sin(c/2)|n\rangle \\
 &=& {\textstyle\frac{1}{2}}(\langle c|n+1\rangle+\langle c|n-1\rangle)-
 {\textstyle\frac{1}{2}}\I\tau_i (\langle c|n+1\rangle-\langle c|n-1\rangle)\;.
\end{eqnarray*}
Thus, in a triad representation holonomies act by changing the label
$n$ in $\psi_n(\phi)$ by $\pm1$ since, e.g.,
\[
 (\sin(c/2)\psi)_n=
 -{\textstyle\frac{1}{2}}\I\sum_n\psi_n(|n+1\rangle-|n-1\rangle)=
 {\textstyle\frac{1}{2}}\I\sum_n(\psi_{n+1}-\psi_{n-1})|n\rangle\;.
\]

The constraint operator contains several holonomy operators and also
the volume operator. It leads to the constraint
equation \cite{IsoCosmo,Closed}
\begin{eqnarray} \label{Evolve}
&& (V_{|n+4|/2}-V_{|n+4|/2-1}) \E^{\I k} \psi_{n+4}(\phi)-
 (2+\gamma^2k^2) (V_{|n|/2}-V_{|n|/2-1}) \psi_{n}(\phi)\nonumber\\
&& + (V_{|n-4|/2}-V_{|n-4|/2-1}) \E^{-\I k}
 \psi_{n-4}(\phi)\nonumber\\
&=& -\frac{8\pi}{3}G\gamma^3\lP^2\hat{H}_{\rm matter}(n)\psi_n(\phi)
\end{eqnarray}
which is a {\em difference equation\/} rather than a differential
equation thanks to the discrete internal time. The parameter $k$ again
signifies the intrinsic curvature; for technical reasons the above
equation has only been derived for the values $k=0$ and $k=1$, not for
$k=-1$.

While the left hand side is very different from the Wheeler--DeWitt
case, the right hand side looks similar. This is, however, only
superficially so; for we have to use the quantizations of the
preceding subsection for inverse metric components, in particular in
the kinetic term.

We can eliminate the phase factors $\E^{\pm\I k}$ in (\ref{Evolve}) by
using a wave function $\tilde{\psi}_n(\phi):=\E^{\I nk/4}\psi_n(\phi)$
which satisfies the same equation without the phase factors (of
course, it is different from the original wave function only for
$k=1$). The phase factor can be thought of as representing rapid
oscillations of the wave function caused by non-zero intrinsic
curvature.

\subsubsection{Large volume behavior}

Since the Wheeler--DeWitt equation corresponds to a straightforward
quantization of the model, it should at least approximately be valid
when we are far away from the singularity, i.e.\ when the volume is
large enough. To check that it is indeed reproduced we assume large
volume, i.e.\ $n\gg 1$, and that the discrete wave function
$\psi_n(\phi)$ does not display rapid oscillations at the Planck
scale, i.e.\ from $n$ to $n+1$, because this would indicate a
significantly quantum behavior. We can thus interpolate the discrete
wave function by a continuous one $\tilde{\psi}(p,\phi)=
\tilde\psi_{n(p)}(\phi)$ with $n(p)=6p/\gamma\lP^2$ from (\ref{p}). By
our assumption of only mild oscillations, $\tilde{\psi}(p,\phi)$ can
be assumed to be smooth with small higher order derivatives. We can
then insert the smooth wave function in (\ref{Evolve}) and perform a
Taylor expansion of $\tilde{\psi}_{n\pm 4}(\phi)=
\tilde{\psi}(p(n)\pm\frac{2}{3} \gamma\lP^2)$ in terms of
$p/\gamma\lP^2$. It is easy to check that this yields to leading order
the equation
\[
 \frac{1}{2}\left(\frac{4}{9}\lP^4\frac{\partial^2}{\partial
 p^2}-k\right) \tilde{\psi}(p,\phi)= -\frac{8\pi}{3}G\hat{H}_{\rm
 matter}(p)\tilde{\psi}(p,\phi)
\]
which with $a=\sqrt{|p|}$ is the Wheeler--DeWitt equation (\ref{WdW})
in the ordering given before \cite{SemiClass}. Thus, indeed, at large
volume the Wheeler--DeWitt equation is reproduced which demonstrates
that the difference equation has the correct continuum limit at large
volume. It also shows that the old Wheeler--DeWitt quantization,
though not the fundamental evolution equation from the point of view
of quantum geometry, can be used reliably as long as only situations
are involved where the discreteness is not important. This includes
many semiclassical situations, but not questions about the singularity.

When the volume is small, we are not allowed to do the Taylor
expansions since $n$ is of the order of one. There we expect important
deviations between the difference equation and the approximate
differential equation. This is close to the classical singularity,
where we want corrections to occur since the Wheeler--DeWitt
quantization cannot deal with the singularity problem.

\subsubsection{Non-singular evolution}

To check the issue of the singularity in loop quantum cosmology we
have to use the exact equation (\ref{Evolve}) without any
approximations \cite{Sing}. We start with initial values for
$\psi_n(\phi)$ at large, positive $n$ where we know that the behavior
is close to the classical one. Then, we can evolve backwards using the
evolution equation as a recurrence relation for $\psi_{n-4}(\phi)$ in
terms of the initial values. In this way, we evolve toward the
classical singularity and we will be able to see what happens
there. The evolution is unproblematic as long as the coefficient
$V_{|n-4|/2}-V_{|n-4|/2-1}$ of $\psi_{n-4}$ in the evolution equation
is non-zero. It is easy to check, however, that it can be zero, if and
only if $n=4$. When $n$ is four, we are about to determine the value
of the wave function at $n=0$, i.e.\ right at the classical
singularity, which is thus impossible. It seems that we are running
into a singularity problem again: the evolution equation does not tell
us the value $\psi_0(\phi)$ there.

A closer look confirms that there is {\em no\/} singularity. Let us
first ignore the values $\psi_0(\phi)$ and try to evolve {\em
through\/} the classical singularity. First there are no problems: for
$\psi_{-1}(\phi)$ we only need $\psi_3(\phi)$ and $\psi_7(\phi)$ which
we know in terms of our initial data. Similarly we can determine
$\psi_{-2}(\phi)$ and $\psi_{-3}(\phi)$. When we come to
$\psi_{-4}(\phi)$ it seems that we would need the unknown
$\psi_0(\phi)$ which, fortunately, is not the case because
$\psi_0(\phi)$ drops out of the evolution equation completely. It does
not appear in the middle term on the left hand side because now, for
$n=0$, $V_{|n|/2}-V_{|n|/2-1}=0$. Furthermore, we have seen as a
general conclusion of loop quantizations that the matter Hamiltonian
annihilates the singular state $|0\rangle$, which in the triad
representation translates to $\hat{H}_{\rm matter}(n=0)=0$ independently
of the kind of matter. Thus, $\psi_0(\phi)$ drops out completely and
$\psi_{-4}$ is determined solely by $\psi_4$. The further evolution to
all negative $n$ then proceeds without encountering any problems.

Intuitively, we obtain a branch of the universe at times ``before''
the classical singularity, which cannot be seen in the classical
description nor in the Wheeler--DeWitt quantization. Note, however,
that the classical space-time picture and the notion of time resolves
around the singularity; the system can only be described by quantum
geometry. The branch at negative times collapses to small volume,
eventually reaching volume zero in the Planck regime. There, however,
the evolution does not stop, but the universe bounces to enter the
branch at positive time we observe. During the bounce, the universe
``turns its inside out'' in the sense that the orientation of space,
given by $\sgn(p)$, changes.

In the discussion we ignored the fact that we could not determine
$\psi_0(\phi)$ by using the evolution equation. Is it problematic that
we do not know the values at the classical singularity? There is no
problem at all because those values just decouple from values at
non-zero $n$. Therefore, we can just choose them freely; they do not
influence the behavior at positive volume. In particular, they cannot
be determined in the above way from the initial data just because they
are completely independent.

The decoupling of $\psi_0(\phi)$ was crucial in the way we evolved
through the classical singularity. Had the values not decoupled
completely, it would have been impossible to continue to all negative
$n$. It could have happened that the lowest order coefficient is zero
at some $n$, not allowing to determine $\psi_{n-4}$, but that this
unknown value would not drop out when trying to determine lower
$\psi_n$. In fact, this would have happened had we chosen a factor
ordering different from the one implicitly assumed above. Thus, the
requirement of a non-singular evolution selects the factor ordering in
loop quantum cosmology which, in turn, fixes the factor ordering of
the Wheeler--DeWitt equation (\ref{WdW}) via the continuum limit. One
can then re-check results of Wheeler--DeWitt quantum cosmology which
are sensitive to the ordering \cite{Konto} with the one we obtain
here. It is not one of the orderings usually used for aesthetic
reasons such that adaptations can be expected. An initial step of the
analysis has been done in \cite{Closed}.

To summarize, the evolution equation of loop quantum cosmology allows
us, for the first time, to push the evolution through the classical
singularity. The theory tells us what happens beyond the classical
singularity which means that there is no singularity at all. We
already know that energy densities do not diverge in a loop
quantization, and now we have seen that the evolution does not
stop. Thus, none of the conditions for a singularity is satisfied.

\subsubsection{Dynamical initial conditions}

In the Wheeler--DeWitt quantization the singularity problem has been
glossed over by imposing initial conditions at $a=0$, which does have
the advantage of selecting a unique state (up to norm) appropriate for
the unique universe we observe. This issue appears now in a new light
because $n=0$ does not correspond to a ``beginning'' so that it does
not make sense to choose initial conditions there. Still, $n=0$ does
play a special role, and in fact the behavior of the evolution
equation at $n=0$ {\em implies\/} conditions for a wave function
\cite{DynIn}. The dynamical law and the issue of initial conditions
are intertwined with each other and not separate as usually in
physics. One object, the constraint equation, both governs the
evolution and provides initial conditions. Due to the intimate
relation with the dynamical law, initial conditions derived in this
way are called {\em dynamical initial conditions}.

To see this we have to look again at the recurrence performed
above. We noted that the constraint equation does not allow us to
determine $\psi_0(\phi)$ from the initial data. We then just ignored
the equation for $n=4$ and went on to determine the values for
negative $n$. The $n=4$-equation, however, is part of the constraint
equation and has to be fulfilled. Since $\psi_0$ drops out, it is a
linear condition between $\psi_4$ and $\psi_8$ or, in a very implicit
way, a linear condition for our initial data. If we only consider the
gravitational part, i.e.\ the dependence on $n$, this is just what we
need. Because the second order Wheeler--DeWitt equation is reproduced
at large volume, we have a two-parameter freedom of choosing the
initial values in such a way that the wave function oscillates only
slowly at large volume. Then, one linear condition is enough to fix
the wave function up to norm. When we also take into account the
matter field, there is still more freedom since the dependence of the
initial value on $\phi$ is not restricted by our condition. But the
freedom is still reduced from two functions to one. Since we have
simply coupled the scalar straightforwardly to gravity, its initial
conditions remain independent. Further restrictions can only be
expected from a more universal description. Note also that there are
solutions with a wave length the size of the Planck length which are
unrestricted (since the evolution equation only relates the wave
function at $n$ and $n\pm 4$). Their role is not understood so far,
and progress can only be achieved after the measurement process or, in
mathematical terms, the issue of the physical inner product is better
understood.

In its spirit, the dynamical initial conditions are very different
from the old proposals since they do not amount to prescribing a value
of the wave function at $a=0$. Still, they can be compared at least at
an approximate level concerning implications for a wave function. They
are quite similar to DeWitt's original proposal that the wave function
vanishes at $a=0$. The value at $a=0$ itself would not be fixed, but
quite generally the wave function has to approach zero when it reaches
$n=0$. In this sense, the dynamical initial conditions can be seen to
provide a generalization of DeWitt's initial condition which does not
lead to ill-posed initial value problems \cite{Scalar}.

For the closed model with $k=1$ we can also compare the implications
with those of the tunnelling and the no-boundary proposals which have
been defined only there. It turns out that the dynamical initial
conditions are very close to the no-boundary proposal while they
differ from the tunnelling one \cite{Closed}.

\subsection{Phenomenology}

So far, we have discussed mainly conceptional issues. Now that we know
that loop quantum cosmology is able to provide a complete,
non-singular description of a quantum universe we can also ask whether
there are observational consequences. We already touched this issue by
comparing with the older boundary proposals which have been argued to
have different implications for the likelihood of inflation from the
initial inflaton values they imply. However, the discussion has not
come to a definite conclusion since the arguments rely on assumptions
about Planck scale physics and also the interpretation of the wave
function about which little is known. Loop quantum cosmology provides
a more complete description and thus can add substantially to this
discussion. An analysis analogous to the one in the Wheeler--DeWitt
quantization has not been undertaken so far; instead a strategy has
been used which works with an effective classical description
implementing important quantum geometry effects and thus allows to
sidestep interpretational problems of the wave function. It provides a
general technique to study quantum effects in a phenomenological way
which is currently being used in a variety of models.

\subsubsection{Effective Friedmann equation}

The central idea is to isolate the most prominent effects of quantum
geometry and transfer them into effective classical equations of
motion, in the case of isotropic cosmology an effective Friedmann
equation \cite{Inflation}. The most prominent effect we have seen is
the cut-off observed for inverse powers of the scale factor (which can
be thought of as a curvature cut-off). It is a non-perturbative effect
and has the additional advantage that its reach can be extended into the
semiclassical regime by choosing a large ambiguity parameter $j$.

\begin{figure}
\centering
 \includegraphics[height=4cm]{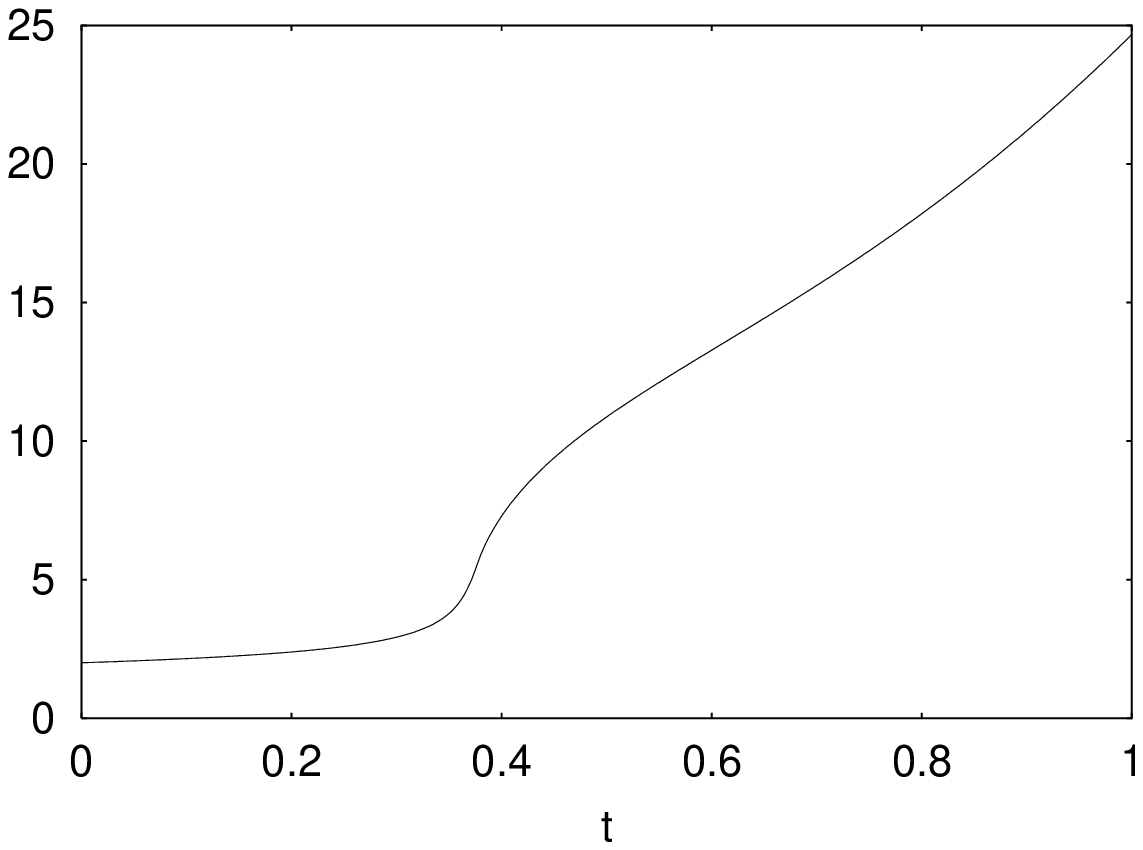}
 \includegraphics[height=4cm]{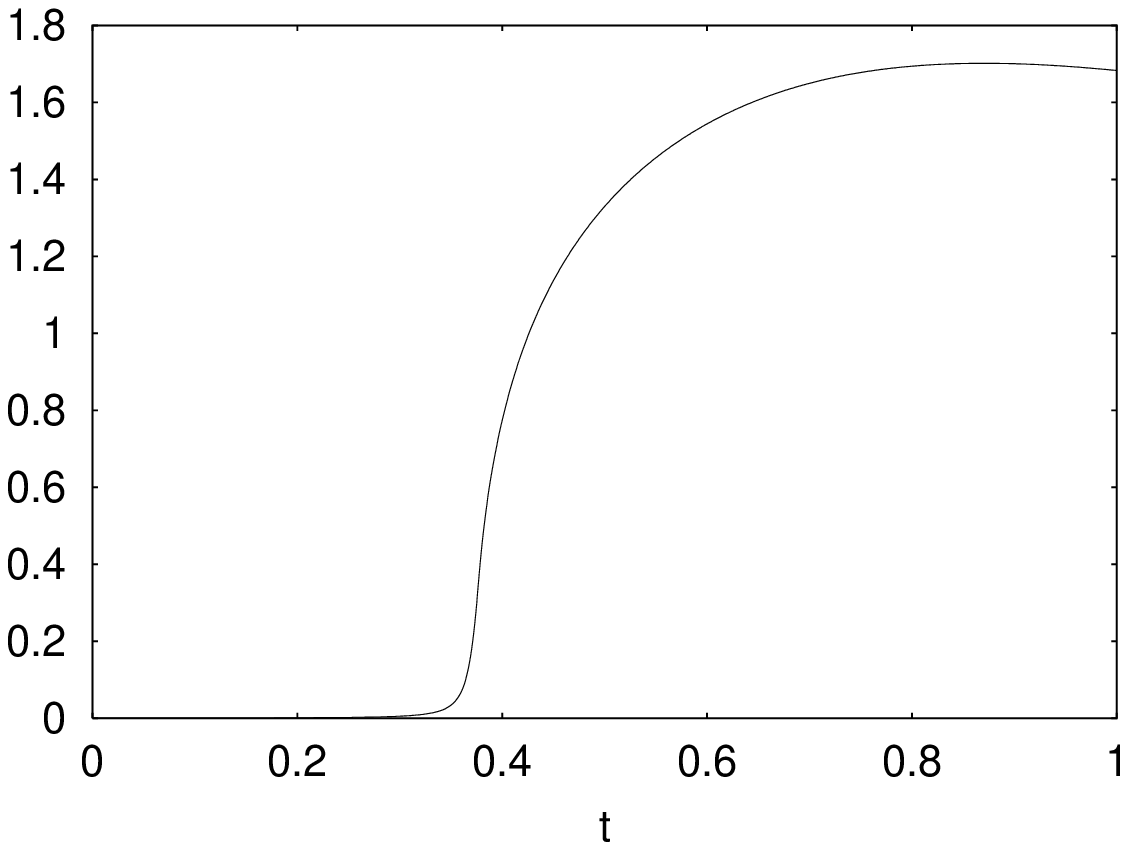}
\caption{Behavior of the scale factor (left) and the scalar field
 (right) during quantum geometry inflation (ending at $t\approx 0.4$
 for $j=100$) \cite{Closed}, both plotted in Planck units. The
 potential is just a mass term $8\pi G W(\phi)=10^{-3}\hbar\phi^2/2$,
 and initial conditions for the numerical integration are $\phi_0=0$,
 $\sqrt{\kappa}\dot{\phi}_0= 10^{-5}\lP^{-1}$ at $a_0=2\lP$.}
\label{SlowRoll}
\end{figure}

In the equations for the isotropic model, inverse powers of the scale
factor appear in the kinetic term of the matter Hamiltonian, e.g.\
(\ref{Hphi}) for a scalar. We have discussed in Section \ref{s:GR}
that it is difficult to suppress this term by arranging the evolution
of $\phi$. Now we know, however, that quantum geometry provides a
different suppression mechanism in the inverse scale factor
operator. This has already played an important role in showing the
absence of singularities since in fact the matter Hamiltonian vanishes
for $n=0$. Instead of $a^{-3}$ we have to use a quantization of the
inverse scale factor, e.g.\ in the form $\widehat{a^{-3}}$ whose
eigenvalues (\ref{ainvm}) with $m=3$ are bounded above. We can
introduce the effect into the classical equations of motion by
replacing $d(a)=a^{-3}$ with the bounded function
$d_j(a)=a^{-3}p(3a^2/j\gamma\lP^2)^6$ where $p(q)$ is defined in
(\ref{pq}) and we choose a half-integer value for $j$. The effective
scalar energy density then can be parameterized as
\[
 \rho_{\rm eff}(a)={\textstyle\frac{1}{2}}x\,a^{l(a)-3}\lP^{-l(a)-3}\, p_{\phi}^2 +
 W(\phi)
\]
with parameters $x$ and $l$ which also depend on quantization
ambiguities, though not significantly. Note that $\lP$ appears in the
denominator demonstrating that the effect is non-perturbative; it
could not be obtained in a perturbative quantization in an expansion
of $G$. With our function $p(q)$ we have $l=12$ for very small $a$
(but it decreases with increasing $a$), and it is usually larger than
$3$, even taking into account different quantization choices, thanks
to the high power of $p(q)$ in $d_j(a)$. Thus, the effective equation
of state parameter $w=-l/3$ in the parametrization (\ref{inflation})
is smaller than $-1$; quantum geometry {\em predicts that the universe
starts with an initial phase of inflation} \cite{Inflation}. It is a
particular realization of super-inflation, but since $w$ increases
with $a$, there is no pole as would be the case with a constant
$w$. Note that this does not require any special arrangements of the
fields and their potentials, not even an introduction of a special
inflaton field: any matter Hamiltonian acquires the modified kinetic
term such that even a vanishing potential implies inflation. Inflation
appears as a natural part of cosmological models in loop quantum
cosmology. Moreover, the inflationary phase ends automatically once
the expanding scale factor reaches the value
$a\approx\sqrt{j\gamma/3}\;\lP$ where the modified density reaches its
peak and starts to decrease (Fig.~\ref{SlowRoll}).

\subsubsection{Inflation}

Thus, inflation appears as a natural consequence, but it is less clear
what role it can play. For an inflationary period responsible for
structure formation it has to last long enough (in terms of
e-foldings, i.e.\ a large ratio of the final $a$ and the initial $a$)
and to be very close to standard inflation, i.e.\ $w=-1$. The final
scale factor for quantum geometry inflation is easy to find,
$a\approx\sqrt{j\gamma/3}\;\lP$ as just discussed. The initial value,
however, is more complicated. In a flat model, the inflationary period
starts as close to $a=0$ as we want, but at those small values the
effective classical description must break down. Moreover, in a closed
model the region close to $a=0$ is classically
forbidden\footnote{There is, however, a mechanism which leads to a
small classically allowed region for small $a$ including $a=0$ even in
a closed model \cite{Closed}. This comes from a suppression of
intrinsic curvature analogous to the cut-off of $a^{-1}$ which would
be a suppression of extrinsic curvature.}  which also sets a lower
limit for the initial $a$. All these issues depend more sensitively on
the kind of matter added to the model and have not yet been analyzed
systematically.

An alternative application of quantum geometry inflation can be seen in
combination with standard inflation. We have discussed that the
standard scenario requires a special potential and also special, very large
initial values for the inflaton. For instance, for chaotic inflation
with the potential $W(\phi)=\frac{1}{2}m\phi^2$ we need to start with
$\phi_0>m_{\rm P}=\hbar/\lP$ which is huge compared to its own mass
$m$. If we couple quantum geometry inflation with chaotic inflation,
we would first observe an inflationary expansion at small volume which
can stop at small $a$ (i.e.\ $j$ can be of the order one). During this
phase also the evolution of the scalar is modified compared to the
standard one since now $d_j(a)$ appears in the Hamiltonian equations
of motion instead of $a^{-3}$. This leads to a differential equation
\[
 \ddot{\phi}=\frac{\D\log d_j(a)}{\D a} \dot{a}\dot{\phi}- a^3d_j(a)
 W'(\phi)
\]
for $\phi$. For the always decreasing $a^{-3}$ instead of $d_j(a)$ we
obtain the previous equation (\ref{phidd}) with the friction term. The
modified $d_j(a)$, however, is increasing for small $a$ such that we
obtain a friction term with the opposite sign. This will require the
inflaton to move up the potential, reaching large values even if it
would start in $\phi(a=0)=0$ \cite{Closed}; see Fig.~\ref{SlowRoll}.

\subsection{Homogeneous cosmology}

The framework of loop quantum cosmology is available for all
homogeneous, but in general anisotropic models \cite{cosmoI}. When we
require that the metric of a homogeneous model is diagonal, the volume
operator simplifies again allowing an explicit analysis
\cite{HomCosmo}. One obtains a more complicated evolution equation
which is now a partial difference equation for three degrees of
freedom, the three diagonal components of the metric. Nevertheless,
the same mechanism for a removal of the classical singularity as in
the isotropic case applies.

This is in particular important since it suggests an absence of
singularities even in the full theory. It has been argued \cite{BKL}
that close to singularities points on a space-like slice decouple from
each other such that the metric in each one is described by a
particular homogeneous model, called Bianchi IX. If this is true and
extends to the quantum theory, it would be enough to have a
non-singular Bianchi IX model for singularity freedom of the full
theory. Even though the classical evolution of the Bianchi IX model is
very complicated and suspected to be chaotic \cite{Chaos}, one can see
that its loop quantization is singularity-free \cite{Spin}. In fact,
again the cut-off in the inverse scale factor leads to modified
effective classical equations of motion which do not show the main
indication for chaos. The Bianchi IX universe would still evolve in a
complicated way, but its behavior simplifies once it reaches small
volume. At this stage, a simple regular transition through the
classical singularity occurs. This issue is currently being
investigated in more detail. Also other homogeneous models provide a
rich class of different systems which can be studied in a
phenomenological way including quantum geometry modifications. 

\section{Quantum gravity phenomenology}
\label{s:Phen}

Since quantum gravity is usually assumed to hold at scales near the Planck length $\lP \sim 10^{-32}$cm or, equivalently, Planck energy $E_{\rm P}:={\hbar}/{\lP}\sim 10^{18}$GeV experiments to probe such a regime were considered out of reach for most of the past. 
Recently, however, phenomena have been proposed which compensate for the tiny size of the Planck scale by a large number of small corrections adding up. These phenomena include in vacuo dispersion relations for gamma ray astrophysics \cite{HUET,AC,gleiser}, laser-interferometric limits on distance fluctuations \cite{ACLFLUC,NGVDAMLFLUC}, neutrino oscillations \cite{BRUSTEIN}, threshold shifts in ultra high energy cosmic ray physics \cite{KIFUNE,ACPIRAN,ALF,MAJOR}, CPT violation \cite{CPTELLIS} and clock-comparison experiments in atomic physics \cite{SUDVUUR}. 
They form the so called {\em quantum gravity phenomenology} \cite{QGPhen}. 

The aim is to understand the imprint which the structure of space-time predicted by a specific theory of quantum gravity can have on matter propagation.
Specifically, dispersion relations are expected to change due to a non-trivial microscopic structure (as in condensed matter physics where the dispersion relations deviate from the continuum approximation once atomic scales are reached). For particles with energy $E\ll E_{\rm P}$  and momentum $\vec p\,$  the following modified vacuum dispersion relations have been proposed \cite{AC}:
\begin{eqnarray}
{\vec p}^2 &=& E^2\left(1 + \xi\, E/E_{\rm P}+ {\cal O}\left(
(E/E_{\rm P})^2\right) \right)\,,\quad  \label{eq:drel} 
\end{eqnarray}
where $\xi\sim 1$ has been assumed, which still has to be verified in concrete realizations. Furthermore, this formula is based on a power series expansion which rests on the assumption that the momentum is analytic at $E=0$ as a function of the energy. In general, the leading corrections can behave as $\left(E/E_{\rm P}\right)^{\Upsilon+1}$, where ${\Upsilon}\geq 0$ is a positive real number.

Since the Planck energy is so large compared to that of particles which can be observed from Earth, the correction would be very tiny even if it is only of linear order. However, if a particle with the modified dispersion relation travels a long distance, the effects can become noticeable. For instance, while all photons travelling at the speed of light in Minkowski space would arrive at the same time if they had been emitted in a brief burst, Eq.\ (\ref{eq:drel}) implies an energy dependent speed for particles with the modified dispersion relations. Compared to a photon travelling a distance $L$ in Minkowski space,
the retardation time is 
\begin{eqnarray}
\Delta t \approx \xi\, L\, E/E_{\rm P}\,.
\label{eq:Deltat}
\end{eqnarray}
If $L$ is of a cosmological scale, the smallness of $E/E_{\rm P}$ can be compensated, thus bringing $\Delta t$ close to possible observations. Candidates for suitable signals are
Gamma Ray Bursts (GRB's), intense short bursts of energy around $E\sim 0.20$MeV that travel a cosmological distance $L\sim 10^{10}\,$ly until they reach Earth.
These values give $\Delta t\sim 0.01$ms which is only two orders of magnitude below the sensitivity $\delta t$ for current observations of GRB's \cite{METZ,BHAT} (for planned improvements see \cite{meszaros}). 
For the delay of two photons detected with an energy difference $\Delta E$, the observational  bound  $E_{\rm P}/\xi\geq 4\times 10^{16}$ GeV was established in \cite{BILLER} by identifying events having $\Delta E=1$ TeV arriving to Earth from the active galaxy Markarian 421 within the time resolution $\Delta t =280\,$s of the measurement. 
Moreover, GRB's also seem to generate Neutrino Bursts (NB) in the range $10^{5}-10^{10}$ GeV in the so-called fireball model \cite{WAX,VIETRI} which can be used for additional observations \cite{ROY,BRUSTEIN,SOUTHHAMPTON}.

In summary, astrophysical observations of photons, neutrinos and also cosmic rays could make tests of quantum gravity effects possible, or at least restrict possible parameters in quantum gravity theories. 

Within loop quantum gravity attention has focused on light  \cite{GRB,Correct2} and neutrino propagation \cite{Correct1}. Other approaches aimed at investigating similar quantum gravity effects include string theory \cite{REVIEW}, an open system approach \cite{opensystems}, perturbative quantum gravity
\cite{oneloop-effqg,PADDY} and non commutative geometry \cite{NC-Amelino}. 
A common feature to all these approaches is that correction terms arise which break Lorentz symmetry.  These studies overlap with a systematic analysis providing a general power counting renormalizable extension of the standard model that incorporates both Lorentz and CPT violations \cite{colladay}. Progress in setting bounds to such symmetry violation has been reported in \cite{gleiser,SUDVUUR,tritium,liberati,myers}.

\subsection{An implementation in loop quantum gravity}

In order to implement the central idea, one needs states
approximating a classical geometry at lengths much larger than the Planck length. The first proposed states of this type in loop quantum gravity were weave states \cite{weave}.
Flat weave states $|W\rangle$ with characteristic length ${\cal L}$ were constructed
as in Section \ref{s:QG} using collections of circles of Planck size radius (measured with the classical background geometry to be approximated) in random orientation. At
distances $d \gg {\cal L}$ the continuous flat classical geometry is reproduced, while
for distances $d \ll {\cal L}$ the discrete structure of space is manifest.
The search for more realistic coherent states, which not only approximate the classical metric, but also its conjugate, the extrinsic curvature, is still ongoing \cite{GCSI,CohState,Fock,FockGrav}.
Current calculations have been done at a heuristic level by assuming simple properties of semiclassical states. The prize to pay is that some parameters, and even their order of magnitude, remain undetermined. Thus, these studies explore possible quantum gravity effects without using details of specific semiclassical states. Once properties of semiclassical states become better known, one can then check if the existing calculations have to be modified.

\begin{figure}
\centering
 \includegraphics[height=4cm]{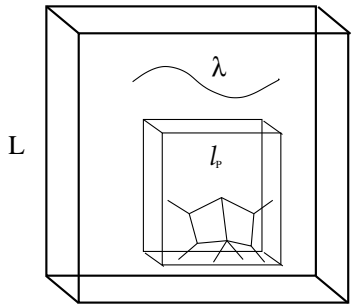}
\caption{Representation of the different scales of the problem. A coarse graining scale 
$\cal L\sim \lambda$ indicates Planck length features $\lP$ are minute as compared to matter scales. $L$ here represents the size of a large piece of space. }
\label{coarse}
\end{figure}

The setup requires to consider semiclassical states $|S\rangle$ for both gravity, the background for the propagation, and the propagating matter.
One has to require that they are peaked at the classical configurations of interest with well defined expectation values such that there exists a coarse-grained expansion involving ratios of the relevant scales of the problem. Those are the
Planck length $\lP$, the characteristic length ${\cal L}$ and the matter
wavelength $\lambda$ satisfying $\lP\ll{\cal L}\leq\lambda$; see Fig.\ \ref{coarse}. The effective Hamiltonian is thus defined by
\begin{equation}
H_{\rm Matter}^{\rm eff}:=<S|{\hat{H}_{\rm Matter}}|S> \,.
\end{equation}

\subsubsection{Light}

The full quantum Hamiltonian for the electromagnetic field is of the form (\ref{HamEM}). With the above assumptions about semiclassical states one can arrive at
an effective electromagnetic Hamiltonian \cite{Correct2}
\ba \label{HEMFIN} H^{\rm eff}_{\rm EM}&=& \frac{1}{Q^2}\int \D^3{\vec x} \left[{\textstyle\frac{1}{2}}\left(1+
\theta_7 \,\left(\lP/{\cal L}\right)^{2+2\Upsilon}
\right)\left(\frac{}{} \underline{{\vec B}}^2 + \underline{{\vec
E}}^2\right)\right.\nonumber\\ && 
+ \left.\theta_3 \, \lP^2 \, \left( \frac{}{}\underline{\vec B}\cdot\nabla^2
\underline{\vec B} + \underline{\vec E}\cdot\nabla^2 \underline{\vec E}\right)
\ \right. \nonumber \\
&&\left. + \theta_8 \lP \Bigl( \underline{\vec B}\cdot(\nabla \times\underline{\vec B})+ 
\underline{\vec E}\cdot(\nabla \times\underline{\vec E}) \Bigr)+\cdots \right], 
\ea
up to order $\lP^2$ and neglecting non linear terms. The coefficients $\theta_i$ have not yet been derived systematically; rather, the expression is to be understood as a collection of all the terms which can be expected. Precise values, and even the order of magnitude, can depend significantly on the explicit procedure followed to obtain the values from a semiclassical state. Moreover, as usual there are quantization ambiguities in the quantum Hamiltonian which influence the coefficients of the correction terms \cite{Ambig}.

From the  effective Hamiltonian
(\ref{HEMFIN}) we obtain the equations of motion
\begin{eqnarray}
A (\nabla\times\underline{\vec B})-\frac{\partial\underline{\vec E}}{\partial t}%
+ 2\lP^2\theta_3\nabla^2(\nabla\times\underline{\vec B}) -2\theta_8\lP\nabla^2\underline{\vec
B} =0,\label{ME1} \\
A (\nabla\times\underline{\vec E})+\frac{\partial\underline{\vec B}}{\partial t}+
2\lP^2\theta_3\nabla^2(\nabla\times\underline{\vec E})-2\theta_8\lP\nabla^2\underline{\vec E}=0,
\label{ME2}
\end{eqnarray}
where
\begin{eqnarray}
A=1+\theta_7\left(\lP/{\cal L}\right)^{2+2\Upsilon}.
\end{eqnarray}
The above equations are supplemented by  $\nabla\cdot \underline{\vec B}=0$, together with the constraint $\nabla\cdot\underline{\vec E}=0$, appropriate for vacuum.

Modifications in the Maxwell equations (\ref{ME1}) and (\ref{ME2}) imply a modified dispersion relation which, neglecting the non-linear part, can be derived by introducing the plane wave ansatz
\begin{eqnarray}
\vec E=\vec E_0\,\E^{\I(\vec k\vec x-\omega t)},\qquad \vec B=\vec B_0\, \E^{\I(\vec
k\vec x-\omega t)},\qquad k=|\vec k |\;.
\end{eqnarray}
The result is
\begin{eqnarray}
\omega=k\left(1+\theta_7\left(\lP/{\cal L}\right)^{2+2\Upsilon}-2\,\theta_3\,(k\lP)^2\pm 2\theta_8\,(k\lP ) \right)
\end{eqnarray}
where the two signs of the last term correspond to the different polarizations of the photon. The speed of a photon becomes
\begin{eqnarray}
v=\frac{\D\omega}{\D k}\left|\frac{}{}\right._{{\cal L}%
=1/k}=1 \pm 4\, \theta_8 \,(k \lP)  -6\theta_3 (k \lP)^2+\theta_7\,\left(k{\lP}\right)^{2+2\Upsilon}+\cdots \label{PHOTV}
\end{eqnarray}
The scale ${\cal L}$ has been estimated by its maximal value $1/k$. Clearly (\ref%
{PHOTV}) is valid only for momenta satisfying $(\lP\,k) \ll 1$.

There are also possible non-linear terms in the effective Maxwell equations \cite{Correct2}.
They can become significant in strong magnetic fields, but
the corrections obtained in the corresponding refraction indices are much
smaller than similar effects in Quantum Electrodynamics. Nevertheless, quantum
gravity corrections have distinct signatures: a main difference is that the speed of
photons with polarization parallel to the plane formed by the background magnetic
field and the direction of the wave is isotropic.

\subsubsection{Spin-1/2 particles}

Similarly, one can derive an effective Hamiltonian for a spin-$\frac{1}{2}$ field of mass $m$ \cite{Correct1}: \begin{eqnarray}  \label{EFFHF}
H^{\rm eff}_{{\rm spin}\;\frac{1}{2}}&=& \i\int \D^3 x \left\{
\pi(\vec x) \tau^d\partial_d \ {\hat A} \right. \xi({\vec x})
+ {\rm c.c.}
+(4{\cal L})^{-1} \ \pi({\vec x}) \,{\hat C}\,
\xi({\vec x}) \\
&+& \frac{m}{2\hbar} [\, \xi^T({\vec x})\ \sigma_2
\left( \alpha + \beta \lP \tau^a
\partial_a \right)\xi({\vec x})  
+\pi^T({\vec x}) \left( \alpha  + \beta \lP  \tau^a \partial_a
\right) \sigma_2 \pi({\vec x})] \Bigr\} \nonumber,
\end{eqnarray}
where 
\begin{eqnarray}\label{EFFHF1}
&&{\hat A}=\left(1 + { \kappa}_{1} \left(\lP/{\cal L}\right)^{\Upsilon+1}+ { \kappa}%
_{2} \left(\lP/{\cal L}\right)^{2\Upsilon+2} + {\textstyle\frac{1}{2}}\,\kappa_3 \
\lP^2 \ \ \nabla^2 \right),\nonumber\\ 
&&{\hat C}=
{\textstyle\frac{1}{2}}\,
\kappa_{7} \left(\lP/{\cal L}\right)^{\Upsilon}\ \lP^2 \ \ \nabla^2
\nonumber \\
&&  \alpha= 1 + { \kappa}_{8} 
\left(\lP/{\cal L}\right)^{\Upsilon+1}\qquad,
 \qquad \beta=\kappa_9 + \kappa_{11}\left(\lP/{\cal L} \right)^{\Upsilon+1},
\end{eqnarray}

This leads to wave equations
\begin{eqnarray}  \label{E2}
&&\I\hbar\left[\frac{\partial}{\partial t}-{\hat A }\ {\vec
\sigma} \cdot \nabla -\I\frac{{\hat C}}{2{\cal L}} \right]\xi(t,{\vec x}) +m
\left( \alpha -{\textstyle\frac{1}{2}}\I\beta \lP \ {\vec \sigma}\cdot \nabla \right)\chi(t,{%
\vec x})=0   \\
&&\I\hbar\left[\frac{\partial}{\partial t}+ {\hat A} \ {\vec
\sigma} \cdot \nabla +\I\frac{{\hat C}}{2{\cal L}} \right]\chi(t,{\vec x})
+m\left( \alpha - {\textstyle\frac{1}{2}}\I\beta \lP \ {\vec \sigma}\cdot \nabla \right) \xi(t,{
\vec x})=0  
\label{E21} 
\end{eqnarray}
with $\chi(t,{\vec x})=\I\ \sigma_2 \xi^*(t,{\vec x})$. As before, the dispersion relation can be obtained by inserting plane wave solutions, this time positive and negative energy solutions 
\begin{equation}  \label{PWS}
W({\vec p},h) \E^{\mp\I E t/\hbar \pm \I {\vec p}\cdot {\vec x%
}/\hbar}
\end{equation}
where  $W({\vec p},h)$ are
helicity $({\vec \sigma}\cdot {\hat p})$ eigenstates, with $h=\pm 1$, so
that 
\begin{equation}
W({\vec p},1)=\left(%
\begin{array}{c}
{\rm cos}(\theta/2) \\ 
\E^{\I \phi} \ {\rm sin}(\theta/2)%
\end{array}%
\right), \quad W({\vec p},-1)=\left(%
\begin{array}{c}
-\E^{-\I \phi} \ {\rm sin}(\theta/2) \\ 
{\rm cos}(\theta/2)%
\end{array}
\right).
\end{equation}
For ultra-relativistic neutrinos ($p\gg m$) one obtains
\begin{eqnarray}
\lP E_\pm(p,{\cal L})&=&p\lP +\lP m^2/2p \pm {\textstyle\frac{1}{2}}\left( \lP m\right) ^{2}\kappa _{9} -%
{\textstyle\frac{1}{2}}\,\kappa _{3}\left( \lP p\right) ^{3} \\
&&+\left( \lP/{\cal L}\right)^{\Upsilon +1}\left[ \kappa
_{1}p\lP\mp {\textstyle\frac{1}{4}}\,\kappa _{7}
\left( \lP p\right)^{2}\right]+\left( \lP/{\cal L}\right)^{2\Upsilon +2}\kappa _{2}\, p\lP\nonumber
\label{gen1}
\end{eqnarray}
and
\[
v_\pm(p,{\cal L}) =1-\frac{m^{2}}{2p^{2}}-{\textstyle\frac{3}{2}}\,\kappa _{3}\left( \lP p\right) ^{2}+\left( \lP/{\cal L}\right)^{\Upsilon +1}\left( \kappa
_{1} \mp {\textstyle\frac{1}{2}}\,\kappa _7\lP p \right) +\left( \lP/{\cal L}\right)^{2\Upsilon +2}\, \kappa_{2}.
\]

There are two physically interesting effects related to the dispersion relations just described for neutrinos. Namely neutrino oscillations for different flavors and time delay between neutrinos and photons coming from the same GRB. Estimates in this respect have been obtained in \cite{Correct1}.

\subsection{Summary}

The phenomenological considerations described here are intended to give an idea of possible consequences of quantum gravity corrections.
They start with assumptions about a state approximating a classical flat
metric, a classical flat gravitational connection and a generic classical
matter field, at scales larger than the coarse-grained characteristic length
${\cal L}\gg\lP$. Under these assumptions, modified dispersion relations can be expanded in the Planck length.
In general, there are different types of corrections, which can have different dependence on, e.g., the helicity or the scale $\cal L$. This also includes the parameter $\Upsilon$ encoding our (current) ignorance of the scaling of the gravitational connection in a semiclassical expectation value.

The following motivations for the value of $\Upsilon$ have been made \cite{Correct2}:
(i) $\Upsilon=0$ can be understood as that the connection can not be probed below the
coarse graining scale $\cal L$. The corresponding correction scales as $(k\lP)^2$.
(ii) $\Upsilon=1$ may be interpreted as the analog of
a simple analysis \cite{HEITLER}, based on a saturation of Heisenberg's uncertainty relation
inside a box of volume ${\cal L}^3$: $\Delta E \sim \lP/\cal L$, $\Delta
A\sim \lP/{\cal L}^2$ and $\Delta E \Delta A \sim G \hbar/{\cal
L}^3$. Then the correction behaves as $(k\lP)^4$. (iii) A value
$\Upsilon=-\frac{1}{2}$ would lead to a helicity independent first order
correction (i.e. $(k\lP)$); a negative value, however, is not allowed. Further fractional values have been obtained in \cite{QFTonCST} from a detailed proposal for coherent states in loop quantum gravity \cite{GCSI,CohState}.

From an observational point of view, lower order correction terms would certainly be preferable. Most of the evaluations so far have been done for first order terms, but recently also higher order corrections have been started to be compared with observations \cite{ALF,higher}.

\section{Outlook}

As discussed in this article, loop quantum gravity is at a stage where physical results are beginning to emerge which will eventually be confronted by observations. To obtain these results, as usually, approximation schemes have to be employed which capture the physically significant contributions of a full theory. In our applications we used the minisuperspace approximation to study cosmological models and a semiclassical approximation for the propagation of particles. We have to stress, however, that these approximation schemes are currently realized at different levels of precision, both having open issues to be filled in. Loop quantum cosmological models are based on symmetric states which have been explicitly constructed as distributional states in the full theory. There are no further assumptions besides the central one of symmetries. A partially open issue is the relation of symmetric operators to those of the full theory. A precise derivation of this relation will complete our understanding of the models and also of the full theory, but it is not expected to imply changes of the physical results since we know that they are robust under quantization ambiguities.

As for loop quantum gravity phenomenology its central ingredient are semiclassical states which are being investigated with different strategies leading to different proposals. The present explicit calculations are based on simple assumptions about semiclassical states which have to be probed in an eventual realization. 
Thus, there is not only a central simplifying assumption, semiclassicality, but also additional assumptions about its realization. These assumptions affect the presence as well as the magnitude of possible correction terms. It is not just the relation between phenomenology and observations, but also the one between phenomenology and the basic theory which has to be understood better.

Furthermore, there are important conceptual issues which are not yet completely understood. For instance, it would be essential to see the emergence of a classical space-time from semiclassical quantum states in order to study a particle moving in a state which approximates Minkowski space. A related issue is the fact that the discreteness of quantum geometry is supposed to lead to correction terms violating Lorentz symmetry. Such a violation, in turn, implies the existence of a distinguished time-like vector. An open conceptual issue is how such a distinguished vector can arise from the discrete formulation.\footnote{Models to understand modifications of the usual Lorentz symmetry have been developed in \cite{SMOLIN,jacobson}.} For this purpose one would need a distinguished rest frame which could be identified using the cosmic background radiation \cite{GRB}.

Future work will progress along several lines according to the different open problems. First, 
at a basic level, the conceptual issues will have to be understood better. In the case of quantum gravity phenomenology this will come as a consequence of additional insights into semiclassical states which are under investigation \cite{GCSI,CohState,Fock,FockGrav}. This will also change the way how explicit calculations are implemented, and the precision of known results will be enhanced leading to a stronger confrontation with observations. Finally, there are many phenomenological effects which have not yet been investigated in the context of loop quantum gravity. Loop effects will lead to changes whose significance regarding observations has to be studied. 

Already the present stage of developments proves that loop quantum gravity is a viable
description of aspects of the real world. It offers natural solutions to problems, as e.g.\ the singularity problem, which in some cases have been open for decades and plagued all other theories developed so far. At the same time, sometimes surprising consequences emerged which lead to a coherent picture of a universe described by a discrete geometry. All this establishes the viability of loop quantum gravity, and we are beginning to test the theory also observationally.

\section*{Acknowledgements}

We thank J.\ Alfaro, A.\ Ashtekar,  H.\ Sahlmann and L.F.\ Urrutia for discussions.
M.B.\ is grateful to the Universidad Aut\'onoma Metropolitana Iztapalapa for hospitality during the completion of this article. The work of M.B.\ was supported in part by NSF grant PHY00-90091 and
the Eberly research funds of Penn State as well as 40745-F CONACyT grant. H.A.M.T.\ acknowledges partial support from 40745-F CONACyT grant.

\end{document}